%% file: tuc3_v2.tex
\documentclass[iop]{emulateapj}
\usepackage{aas_macros}

\usepackage{color}
\usepackage[dvipsnames,svgnames]{xcolor}

\usepackage{natbib}

\usepackage{amsmath}
\usepackage{xspace}
\input{commands}

\newcommand{\feh}         {\mbox{[Fe/H]}}
\newcommand{\kms}         {\ensuremath{\km~\second^{-1}}\xspace}
\newcommand{\ha}          {\mbox{H$\alpha$}}

\def\spose#1{\hbox to 0pt{#1\hss}}
\def\lta{\mathrel{\spose{\lower 3pt\hbox{$\mathchar"218$}}
     \raise 2.0pt\hbox{$\mathchar"13C$}}}
\def\gta{\mathrel{\spose{\lower 3pt\hbox{$\mathchar"218$}}
    \raise 2.0pt\hbox{$\mathchar"13E$}}}

\shorttitle{The Low-Mass Milky Way Satellite Tucana~III}
\shortauthors{Simon et al.}
\slugcomment{Submitted to the AAS Journals}

\begin{document}

\title{Nearest Neighbor: The Low-Mass Milky Way Satellite Tucana~III\altaffilmark{*}}

\altaffiltext{*}{This paper includes data gathered with the 6.5 meter
  Magellan Telescopes located at Las Campanas Observatory, Chile.}

\input{authors.tex}

\begin{abstract}
We present Magellan/IMACS spectroscopy of the recently discovered
Milky Way satellite Tucana~III (Tuc~III).  We identify 26 member stars in
Tuc~III, from which we measure a mean radial velocity of $v_{hel} =
-102.3 \pm 0.4~\mbox{(stat.)} \pm 2.0~\mbox{(sys.)}$~\kms, a velocity dispersion
of $0.1^{+0.7}_{-0.1}$~\kms, and a mean metallicity of $\mbox{[Fe/H]} =
-2.42^{+0.07}_{-0.08}$.  The upper limit on the velocity dispersion is $\sigma < 1.5$~\kms\ at 95.5\%\ confidence, and the corresponding upper limit on the mass within the half-light radius of Tuc~III is $9.0 \times 10^{4}$~\msun.  We cannot rule out mass-to-light ratios as large as 240~\msun/\lsun\ for Tuc~III, but much lower mass-to-light ratios that would leave the system baryon-dominated are also allowed.  We measure an upper limit on the metallicity spread of the stars in Tuc~III of 0.19~dex at 95.5\% confidence.  Tuc~III has a smaller metallicity dispersion and likely a smaller velocity dispersion than any known dwarf galaxy, but a larger size and lower surface brightness than any known globular cluster.  Its metallicity is also much lower than those of the clusters with similar luminosity.  We therefore tentatively suggest that Tuc~III is the tidally-stripped remnant of a dark matter-dominated dwarf galaxy, but additional precise velocity and metallicity measurements will be necessary for a definitive classification.
If Tuc~III is indeed a dwarf galaxy, it is one of the closest external galaxies to the Sun.  
Because of its proximity, the most luminous stars in
Tuc~III are quite bright, including one star at $V=15.7$ that is the
brightest known member star of an ultra-faint satellite.
\end{abstract}

\keywords{dark matter; galaxies: dwarf; galaxies: individual
  (Tucana~III); galaxies: stellar content; Local Group; stars:
  abundances}

\section{INTRODUCTION}
\label{intro}

The discovery of Milky Way satellites in the last few years has reached
an extremely rapid pace, with the total number of objects found since
2015 nearly matching the entire previously known population
\citep{bechtol15,koposov15,dw15b,dw16,laevens15,laevens15b,kim15_peg3, kim15_kim2,martin15,kim15_hor2,torrealba16,torrealba16b}.  The photometric
identification of these objects in the Dark Energy Survey (DES) and
other large surveys has far outstripped the spectroscopic follow-up
efforts needed to characterize them.  Spectroscopic analyses are
available for less than a third of the new satellites.
The lack of information about the internal kinematics and chemical
abundances of these systems greatly diminishes our ability to use them
to study dark matter and early galaxy formation.

One of the most intriguing of the recently discovered systems is
Tucana~III \citep{dw15b}.  At a distance of just 25~kpc, Tuc III is among the nearest dwarf galaxies and dwarf galaxy candidates to the Milky Way, slightly closer
than Sagittarius \citep{hamanowicz16} and Draco~II \citep{laevens15b} and perhaps just beyond Segue~1
\citep{belokurov07}.  If its dark matter content is similar to that of
other satellites with stellar masses of $\sim10^{3}$~\msun, such as
Segue~1 and Coma Berenices \citep{sg07,simon11}, it would be a very
promising target for indirect dark matter detection experiments
\citep[e.g.,][]{Ackermann:2015zua,magic16}.  Moreover, \citet{dw15b} showed that
Tuc~III is at the center of a pair of very low surface brightness
linear features extending out to a radius of 2\degr\ (870~pc).  These
streams of stars are likely tidal tails from the pending destruction
of the system, which would make Tuc~III a prototype for the tidal
disruption of the smallest galaxies.

In this paper we present an initial spectroscopic study of Tuc III,
identifying its brightest member stars and measuring its global
properties.  This work will set the stage for more detailed follow-up
efforts, especially focused on the tidal tails, the dynamical state of
the system, and its chemical abundance patterns.  In 
\S\ref{observations} we describe the instrument configuration, target 
selection, observations, and data reduction.  We present our
measurements of the velocities and metallicities of stars in Tuc~III
in \S\ref{measurements}.  In \S\ref{discussion} we derive the physical properties of Tuc~III and consider its nature and origin.  We summarize 
our results and discuss our conclusions in \S\ref{conclusions}.

\section{OBSERVATIONS AND DATA REDUCTION}
\label{observations}

\subsection{Spectrograph Setup and Summary of Observations}

We observed Tuc~III with the IMACS spectrograph \citep{dressler06} on
the Magellan Baade telescope on five nights in July 2015, one night
in October 2015, and four nights in August/September 2016.  We used 
the f/4 camera on IMACS, which provides a
15.4\arcmin\ square field of view for multi-slit spectroscopy.  The f/4 
camera uses a $4 \times 2$ array of $2048 \times 4096$ pixel e2v
CCDs to create an $8192 \times 8192$ mosaic, with the spectral direction 
corresponding to the 2048~pixel axis of each detector.  Our spectra were 
obtained with two gratings, both ruled at 1200 $\ell$/mm
but with different blaze angles.  The new 1200/32.7\degr\ grating acquired for this project reaches a peak throughput (including the
spectrograph and the telescope) of 15.4\%\ from $7800-8800$~\AA, while
the pre-existing 1200/26.7\degr\ grating peaks at 20\%\ at
$6500-7500$~\AA\ but has only 12\%\ throughput at 8500~\AA.  Both gratings
produce spectra at a resolution of $R \approx 11,000$ (for a
0\farcs7\ slit) with a dispersion of 0.19~\AA~pix$^{-1}$.  We
observed with two spectrograph setups, one covering
$7400-9000$~\AA\ targeting the Ca triplet (CaT) absorption lines
(primarily with the 1200/32.7\degr\ grating) and one covering
$6300-7900$~\AA\ targeting \ha\ (primarily with the
1200/26.7\degr\ grating).  In order to cover these entire wavelength
ranges, which include both strong stellar absorption features (CaT and
\ha) and strong telluric absorption (Fraunhofer A-band) to provide a
wavelength reference \citep[e.g.,][]{sohn07,sg07}, the placement of
slits is limited to a $15\arcmin \times 8\arcmin$ portion of the full
field of view.  All observations were obtained with a wide-band $5600-9200$~\AA\ filter to block second-order light.

Because the positions of spectral lines on the detector array are
repeatable at the level of a few pixels rather than a fraction of a
pixel as our velocity measurements require, we acquire calibration
frames during the night.  A typical observing sequence is to take
several exposures of a slitmask lasting $1-1.5$~hr, followed by
exposures of comparison lamps and a flatfield lamp at
the same position.  Then the target is reacquired by taking a short
exposure with the grating removed but the mask left in place.  For observations obtained through 2015 October the comparison lamps were He, Ne, and Ar, but in later observing runs we replaced the He lamps with Kr lamps to provide additional useful calibration lines in the critical $7600-7900$~\AA\ wavelength range.

We observed a total of five slitmasks targeting candidate member stars in Tuc~III
(see \S\ref{sec:targets}).  Mask~1 contained 63 $0\farcs7 \times
5\farcs0$ slits and was observed in the CaT configuration for a
total of 6.5~hr on the nights of 2015 July 16 and 17.  The observing
conditions on those nights were clear, with seeing ranging from
$0\farcs8 - 1\farcs1$.  We also observed Mask~1 in the
\ha\ configuration for a total of 7~hr on 2015 July 18 and 20, with
partially cloudy conditions and seeing generally above 1\arcsec.
Since the \ha\ spectroscopy was not planned as far in advance as would
ideally be the case, we used the same mask rather than one designed
for the required grating angle.  While most targets were unaffected, the spectra from several
slits fell partially off of the detector or overlapped with each
other.  Mask~2
contained 58 slits (including 41 new stars and 17 that were also on
Mask~1) and was designed primarily to target a few bright stars that could not
be placed on Mask~1.  Mask~2 was observed for 1.0~hr on 2015 July 19
in very poor conditions (1\farcs4\ seeing) and for 0.67~hr on 2015 
October 17 in better weather (0\farcs9\ seeing).  During the unusually 
bad seeing on 2015 July 19 we also observed two very bright stars 
(one of which did not fit on either slit mask) with a 0\farcs7-wide 
long-slit for 0.75~hr.  

For the 2016 observing run we designed three additional CaT slitmasks targeting
both member stars identified in 2015 observations and new member candidates.  
Mask~3 contained 64 slits, including most of the brighter confirmed members
and a large sample of fainter candidates on the main sequence.  We observed
Mask~3 for a total of 10.3~hr on the nights of 2016 August 29 and 30 in good 
conditions with seeing mostly between 0\farcs5 and 0\farcs8.  Masks 4 and 5
were designed to observe all remaining bright ($g < 20$) candidates within 
15\arcmin\ (2.5$\times$ the half-light radius) of the center of Tuc~III.  Mask 4 (with 30 stars)
was observed for 1.0~hr on the night of 2016 August 31 in 0\farcs6$-$1\farcs0 
seeing.  Mask 5 (containing 34 stars) was observed for a total of 1.5~hr on 
2016 August 31 and 2016 September 2 in 0\farcs7 seeing.  The observations of 
each mask are summarized in Table~\ref{obstable}.

\input{obstable.tex}

\subsection{Target Selection}
\label{sec:targets}

\begin{figure*}[th!]
\epsscale{1.2}
\plotone{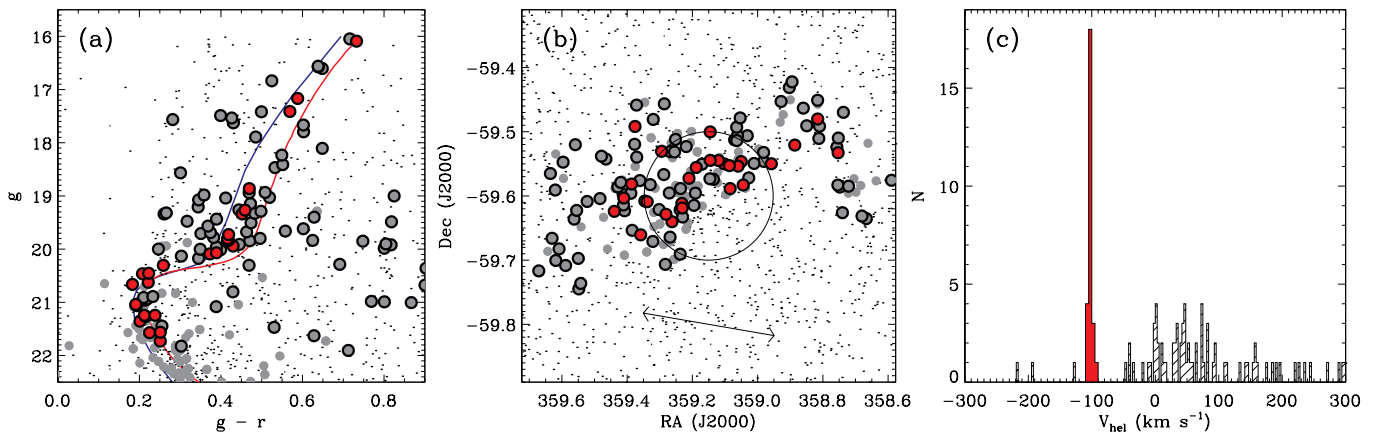}
\caption{\emph{(a)} DES color-magnitude diagram of Tucana~III.  Stars
  within 18\arcmin\ of the center of Tuc~III are plotted as small
  black dots, and stars selected for spectroscopy (as described in
  \S\ref{sec:targets}) are plotted as filled gray circles.  Points
  surrounded by black outlines represent the stars for which we
  obtained successful velocity measurements, and those we identify as
  Tuc~III members are filled in with red.  The M92 sequence and PARSEC
  isochrone used to define the RGB of Tuc~III are displayed as blue
  and red curves, respectively.  \emph{(b)} Spatial distribution of
  the observed stars.  Symbols are as in panel \emph{(a)}.    The arrow at the bottom illustrates the 
  direction of the tidal tails identified by \citet{dw15b}. These tidal tails make it difficult to determine the ellipticity of Tuc~III with existing
  data, so we represent the half-light radius of Tuc~III
  with a black circle. \emph{(c)} Radial velocity distribution of
  observed stars.  The clear narrow peak of stars at $v_{{\rm hel}} \sim
  -100$~\kms highlighted in red is the signature of Tuc~III.  The
  hatched histogram indicates stars that are not members of Tuc~III.}
\label{cmd}
\end{figure*}

We selected targets for spectroscopy using photometry from the DES Year 2 quick release (Y2Q1) catalog \citep{dw15b}.  Guided by the
colors of spectroscopically confirmed members of the DES-discovered
dwarf galaxies Reticulum~II and Horologium~I from \citet{simon15b} and
\citet{koposov15b}, we employed a selection window for red giant
branch (RGB) stars bounded on the blue side by the fiducial sequence
of the metal-poor globular cluster M92 from \citet{an08}, transformed
to the DES photometric system, and bounded on the red side by a
12~Gyr, $\feh = -2.2$ theoretical isochrone from \citet{bressan12}.  
These isochrones are overplotted on the color-magnitude diagram (CMD) 
of Tuc~III in Figure~\ref{cmd}a.  Potential subgiants were selected 
using a 0.05~magnitude-wide window in $g-r$ around the 
\citeauthor{bressan12} isochrone for $20 < g < 20.4$.  We selected 
stars at and below the main sequence turnoff (MSTO) based on the 
photometric membership probabilities derived from the likelihood 
analysis of \citet{dw15b}.  While we did not identify any blue 
horizontal branch (HB) stars likely to be members of Tuc~III,
we also selected a handful of candidate red HB stars at $0.41 < g-r <
0.47$ and $17.4 < g < 17.7$.

Stars were placed on the slitmasks according to their
prioritization in the categories discussed above.  The highest
priority targets were those meeting our RGB selection.  The next
highest priorities were for stars just outside the RGB window, but
within 0.02~mag in $g-r$ of either the M92 or \citet{bressan12}
isochrones.  Subgiants and then main sequence stars with membership
probabilities higher than 0.5 were the next two categories, followed
by main sequence candidates with membership probabilities between 0.1
and 0.5 and red HB candidates.  Finally, any remaining mask space was
filled by stars with photometry that made them unlikely to be members.  Within each category,
priorities were based on brightness and distance from the center of
Tuc~III.

\subsection{Data Reduction}
\label{sec:reductions}

We reduced the IMACS spectra using the following set of procedures.
First, we performed bias subtraction using row-by-row and
column-by-column medians of the column and row bias sections,
respectively.  The f/4 detector array suffers from pattern noise with
a peak-to-peak amplitude of 6 counts that significantly affects the
spectra of the faintest sources.  The pattern varies from one exposure
to the next, but is identical on all eight CCDs for any given frame.
We therefore constructed a template of the pattern by stacking the eight
CCDs and setting each pixel in the template equal to an average of the
two lowest observed values for that pixel.  When there is significant
CCD area that is not covered by spectra this procedure cleanly
reproduced the pattern, but for densely packed data many pixels are covered by a spectrum on all eight CCDs.  In that case, the
template contains artifacts where bright lines or stellar continua
overlap on all of the chips.  We modeled these artifacts by collapsing the template
in the $x$ (dispersion) direction and smoothing it with a Gaussian 
kernel with a full-width at half-maximum of 15 pixels (1\farcs65).  
This smoothed profile was subtracted from each column of the original
template to create the final template, which was then subtracted from the data.

\setcounter{footnote}{36}
The next stage of the data reduction relied on the Cosmos pipeline
described by
\citet{dressler11}.\footnote{http://code.obs.carnegiescience.edu/cosmos}
Cosmos uses the comparison lamp images and the coordinates of the
slits to derive an approximate wavelength solution and a map of
each slit across the detector array.  We then cut out each slit on
each chip into a separate FITS file.

From this point, we adapted the DEEP2 data reduction pipeline
\citep{cooper12, newman13} designed for the DEIMOS spectrograph at
Keck to process IMACS data.  The most significant difference between
DEIMOS and IMACS spectra is that the IMACS spectrum of a single chip
covers a much smaller wavelength range ($\sim390$~\AA\ rather than
$\sim1350$~\AA) because the spectra span the short axis of the
$2048\times4096$ CCDs rather than the long axis and because of the
smaller dispersion.  Deriving a wavelength solution accurate to better
than 1~\kms\ is therefore more challenging because of the small number
of arc lines per chip.  In order to better constrain the problem, we
constructed a single wavelength solution covering all four CCDs for each
slit, fitting the positions of the arc lines with a third-order
Legendre polynomial and leaving the widths of the three chip gaps
spanned by each spectrum as free parameters.  We then had a total of seven parameters (four polynomial coefficients and three chip gaps) to be determined from $\sim35$ arc lines, rather than determining four polynomial coefficients per chip from $\lesssim9$ lines.  With the Cosmos
wavelength solution providing an initial guess for this procedure, we
achieved a typical rms of 0.35~\kms\ for $30-40$ arc lines.  The
remainder of the DEEP2 pipeline, including flatfielding, sky
subtraction, and extraction, required modifications primarily to
handle the different format of the spectra and data files.  All of the
updates to DEIMOS reduction procedures described by \citet{sg07} and
\citet*{kirby15} have been retained in the IMACS version of the
pipeline, with the exception of modifications related to differential
atmospheric refraction because IMACS has an atmospheric dispersion
corrector.

Each individual frame or set of frames was reduced and extracted using
the corresponding calibration exposures.  For the observations of
Masks~1, 3, and 5 we then combined all of the extracted 1D spectra using inverse-variance weighting.  Since the Mask~2 observations were obtained three months apart, radial velocity variations from binary orbital motion are possible \citep[e.g.,][]{koch14}, so we kept the July and October spectra separate.

\section{VELOCITY AND METALLICITY MEASUREMENTS}
\label{measurements}

\subsection{Radial Velocity Measurements}
\label{sec:velocities}

We measured radial velocities from the reduced spectra using the same
procedures described by \citet{sg07} and subsequent papers.  We
observed a set of bright, metal-poor stars in both IMACS
configurations to provide radial velocity template spectra for
$\chi^{2}$ fitting.  We also observed the hot, rapidly rotating star HR~4781 to
serve as a telluric template.  All of the template spectra were
obtained by orienting the IMACS longslit north-south and driving the
telescope in the RA direction such that the star moved at a constant
rate across the slit during the exposure.  These data thus provide a
wavelength reference for a source that uniformly fills the slit.
Integration times for the template observations were set at a minimum
of 120~s so that night sky emission lines would be bright enough to
check the wavelength solution from the comparison lamps.  We reduced
the template spectra as described in \S\ref{sec:reductions}.
The template spectra are normalized, with regions outside of telluric
absorption bands set to unity for the telluric templates and regions
inside the telluric bands set to unity for the velocity templates.

We measured the radial velocity of each science spectrum via
$\chi^{2}$ fits to the velocity templates \citep{sg07,newman13}.  For
this paper we use the metal-poor subgiant HD~140283 \citep{fuhrmann93} as 
the template for all of the science spectra.  We assume a velocity of
$v_{hel} = -171.12$~\kms\ for this star \citep{latham02}.  We
determined a correction to the measured velocities to compensate for possible
mis-centering of each star in its slit by fitting the A-band
absorption of every spectrum with our telluric template.  The A-band
corrections are generally less than 6~\kms, but they show a
systematic dependence on the position of the slit on the mask in the
direction parallel to the slits.  We modeled this dependence as a
quadratic function, and for slits where an accurate correction could
not be measured from the spectrum itself (either because of low S/N or
because the telluric lines landed in a chip gap) we applied the model
correction instead.

As in \citet{sg07}, we calculated statistical uncertainties on each
velocity measurement by performing Monte Carlo simulations in which
normally distributed random noise is added to the observed spectrum
and the template fitting is repeated.  The uncertainty is defined to
be the standard deviation of the Monte Carlo velocity measurements
after $>5\sigma$ outliers have been rejected.  Note that this procedure 
creates a Monte Carlo spectrum that is noisier than the actual data,
and the uncertainty is therefore somewhat overestimated.  For high 
S/N spectra the statistical uncertainty can be as small as 0.1~\kms, 
but that does not mean that the velocity measurements actually achieve 
that absolute accuracy.  The process described above does not take 
into account the uncertainty in the wavelength solution 
($\sim0.35$~\kms), the uncertainty of the A-band correction 
($\sim1$~\kms), uncertainties in the template velocities, and other 
factors.  In order to account for such possible systematic effects, 
we used stars that were observed on multiple nights to quantify the 
overall uncertainties on the velocity measurements.  The systematic uncertainty varies slightly depending on which set of comparison lamps were used to determine the wavelength solution.  For observations obtained with HeNeAr comparison lamps, we analyzed a sample of 62 stars with repeat measurements (including the Tuc~III Mask~1 CaT observations as well as other data using the same spectrograph configuration) and found that we must add a 
systematic uncertainty of 1.2~\kms\ in quadrature with the 
statistical uncertainties to obtain a distribution of velocity 
differences that is consistent with the uncertainties \citep[c.f.][]{sg07,simon15b}.  For observations obtained with KrNeAr lamps the systematic uncertainty determined in the same manner is 1.0~\kms\ because of the improved wavelength solution on the blue end.  
IMACS is therefore able to reach better velocity 
accuracy than Keck/DEIMOS, likely thanks to its higher spectral 
resolution.  We use 1.2~\kms\ (1.0~\kms) as the systematic uncertainty
on the IMACS velocities obtained in 2015 (2016) for the remainder of this paper, with the reported velocity uncertainties being the quadrature sum of the statistical and systematic components.  Combining the uncertainties in this way mitigates the overestimate of the statistical uncertainty mentioned above.  All velocities presented here are in the heliocentric frame.

Because the velocity uncertainties derived from the \ha\ spectra are significantly larger than those determined from the CaT data, we do not use the \ha\ observations for any of the Tuc~III measurements in this paper.  They are used only for display purposes in Fig.~\ref{spectra}.

\subsection{Metallicity Measurements}

We measured metallicities for the red giant members of
Tuc~III based on the equivalent widths of the Ca triplet absorption
lines.  We follow the procedures described in \citet{simon15b} and use
the CaT calibration of \citet{carrera13}.  As was the case for
Reticulum~II, Tuc~III lacks known horizontal branch stars, so we rely
on the CaT calibration with absolute $V$ magnitude rather than the
magnitude relative to the horizontal branch.
The systematic uncertainty on the summed equivalent widths of the CaT lines, determined using repeat measurements in the same way as for the systematic uncertainty on the velocities in \S\ref{sec:velocities}, is 0.32~\AA.

\subsection{Spectroscopic Membership Determination}
\label{membership}

The color-magnitude diagram, spatial distribution, and velocity
distribution of the observed stars are displayed in Figure~\ref{cmd}.
From the sample of stars with reliable velocity measurements (as determined by examination of the $\chi^{2}$ fits to each spectrum), it is evident that
Tuc~III consists of stars in a narrow velocity range from $v_{hel} =
-95$~\kms\ to $v_{hel} = -110$~\kms.  The 26 stars selected with that
velocity cut all lie exactly along the expected RGB-main sequence track of Tuc~III, and all have the spectra of old,
metal-poor stars.  We do not find any other stars within 25~\kms\ of
this velocity, indicating that more sophisticated statistical methods
to identify member stars are not necessary in this data set.  We
classify these 26 objects as Tuc~III members, and all other stars for
which we are able to measure a velocity as non-members.  Velocity
measurements for all stars are listed in Table~\ref{tab:tuc3_spec}.

\begin{figure*}[th!]
\epsscale{1.2}
\plotone{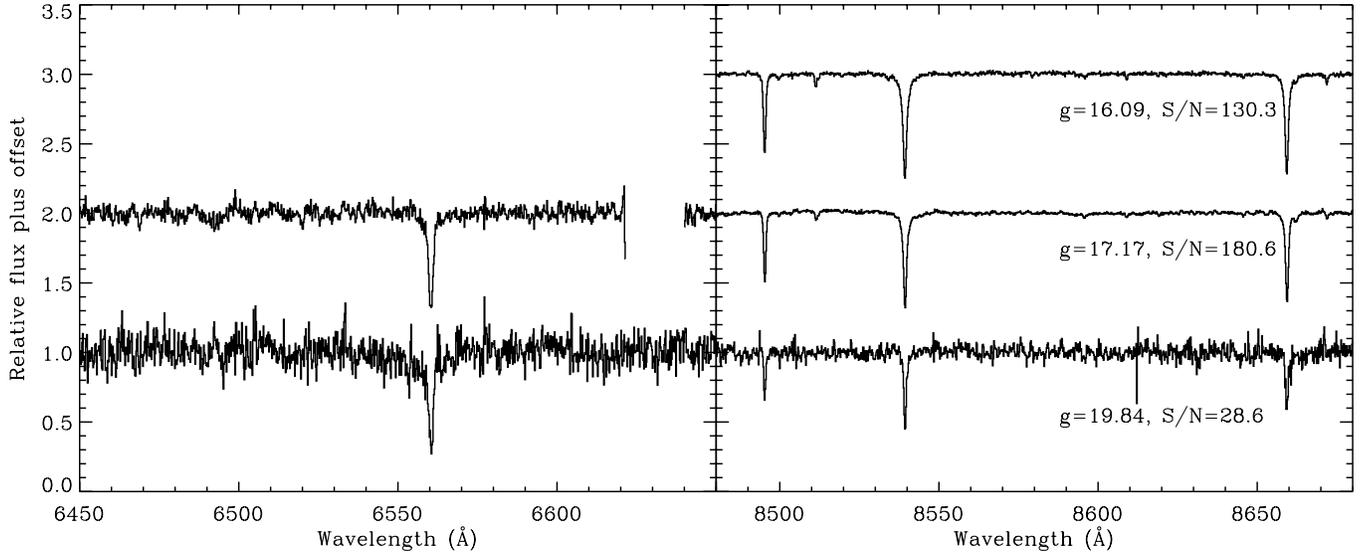}
\caption{IMACS spectra of three Tuc~III RGB stars.  The \ha\ region 
of the spectrum is shown in the left panel, and the CaT region in the 
right panel.  The top spectrum is DES~J235532.66-593114.9, the 
brightest star in Tuc~III, the middle spectrum is 
DES~J235738.48-593611.6, a star $\sim1$~mag fainter, and the 
bottom spectrum is DES~J235655.47-593707.5, near the base of the RGB.  The $\sim20$~\AA\ region with no data visible in the \ha\ spectrum of DES~J235738.48-593611.6 is a gap between IMACS CCDs.  
Note that we did not obtain an \ha\ spectrum of DES~J235532.66-593114.9.}
\label{spectra}
\end{figure*}

Despite its very low luminosity, Tuc~III contains at least three
relatively bright stars.  The most luminous giant we have identified,
DES~J235532.66-593114.9, has a $V$ magnitude of 15.7, $\sim0.3$~mag
brighter than the brightest previously known star in an ultra-faint
dwarf galaxy \citep{simon15b}.  This star and the next two brightest
Tuc~III members (which are both around 17th magnitude) will be
outstanding targets for future chemical abundance studies based on
high resolution spectroscopy.  Example spectra of three Tuc~III members are displayed in Figure~\ref{spectra}.

\section{DISCUSSION}
\label{discussion}

In this section we determine the global properties of Tuc~III, and
then consider their implications for its nature and origin.

\subsection{Velocity Dispersion and Mass}
\label{sec:kinematics}

We determine the mean velocity and velocity dispersion of the Tuc~III
member stars using the Markov Chain Monte Carlo (MCMC) code {\tt emcee}\footnote{http://dan.iel.fm/emcee/current/} 
\citep{2013PASP..125..306F} to maximize the
Gaussian likelihood function defined by \citet{walker06}.  
We find a systemic velocity of $v_{hel} = -102.3 \pm 0.4~\mbox{(stat.)} \pm 2.0~\mbox{(sys.)}$~\kms, where the systematic uncertainty on the mean velocity results from the uncertainty on the velocity zero-point of the template star.  We measure a velocity dispersion of $\sigma =   0.1^{+0.7}_{-0.1}$~\kms.
The 90\%, 95.5\%, and 99.7\%\ upper limits on the velocity dispersion are 1.2~\kms, 1.5~\kms, and 2.3~\kms, respectively.  The only other known Milky Way satellite to have a comparably small dispersion is Segue~2 \citep{kirby13a}, for which the 90\%\ (95\%) upper limit is 2.2~\kms\ (2.6~\kms).

\begin{figure*}[th!]
\epsscale{1.17}
\plottwo{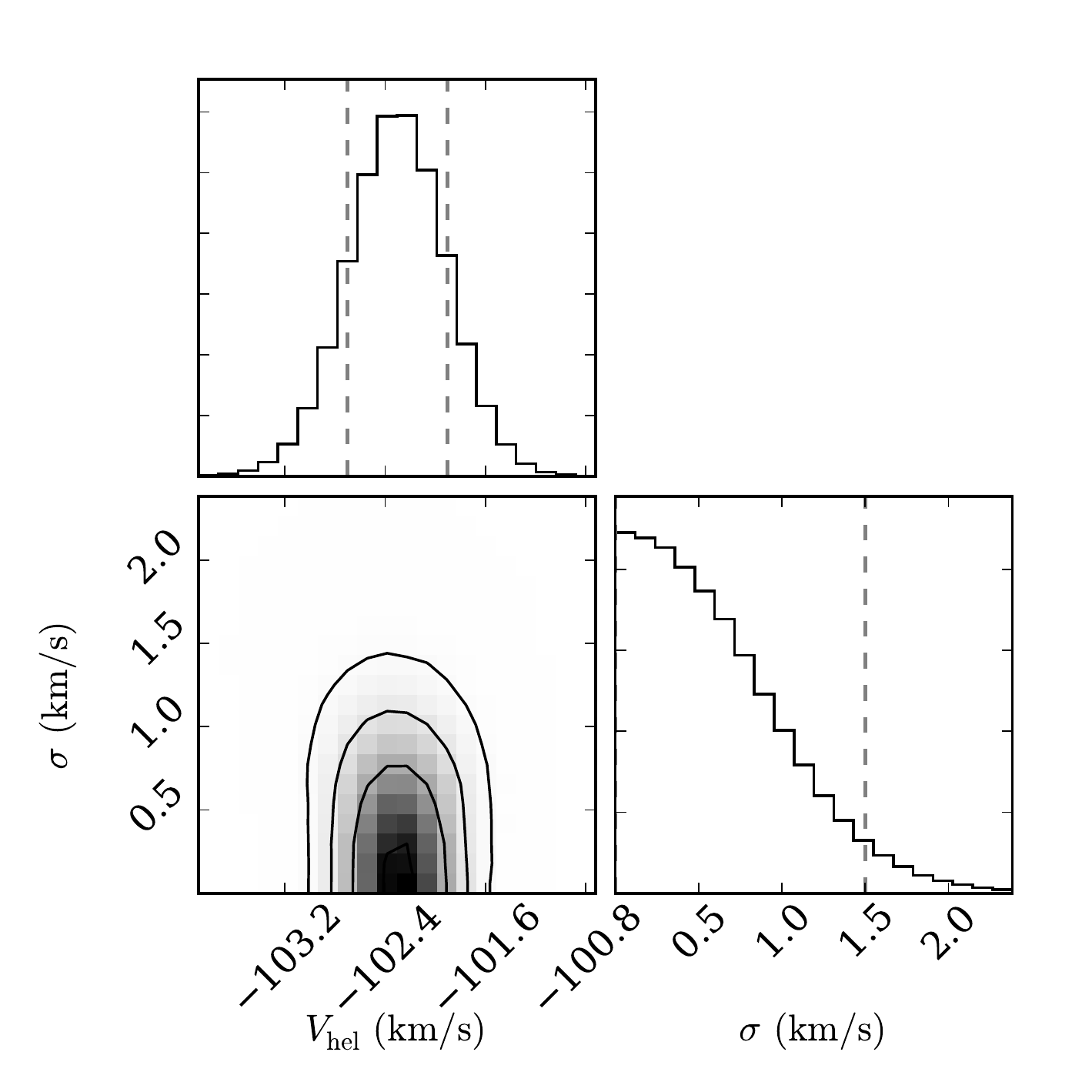}{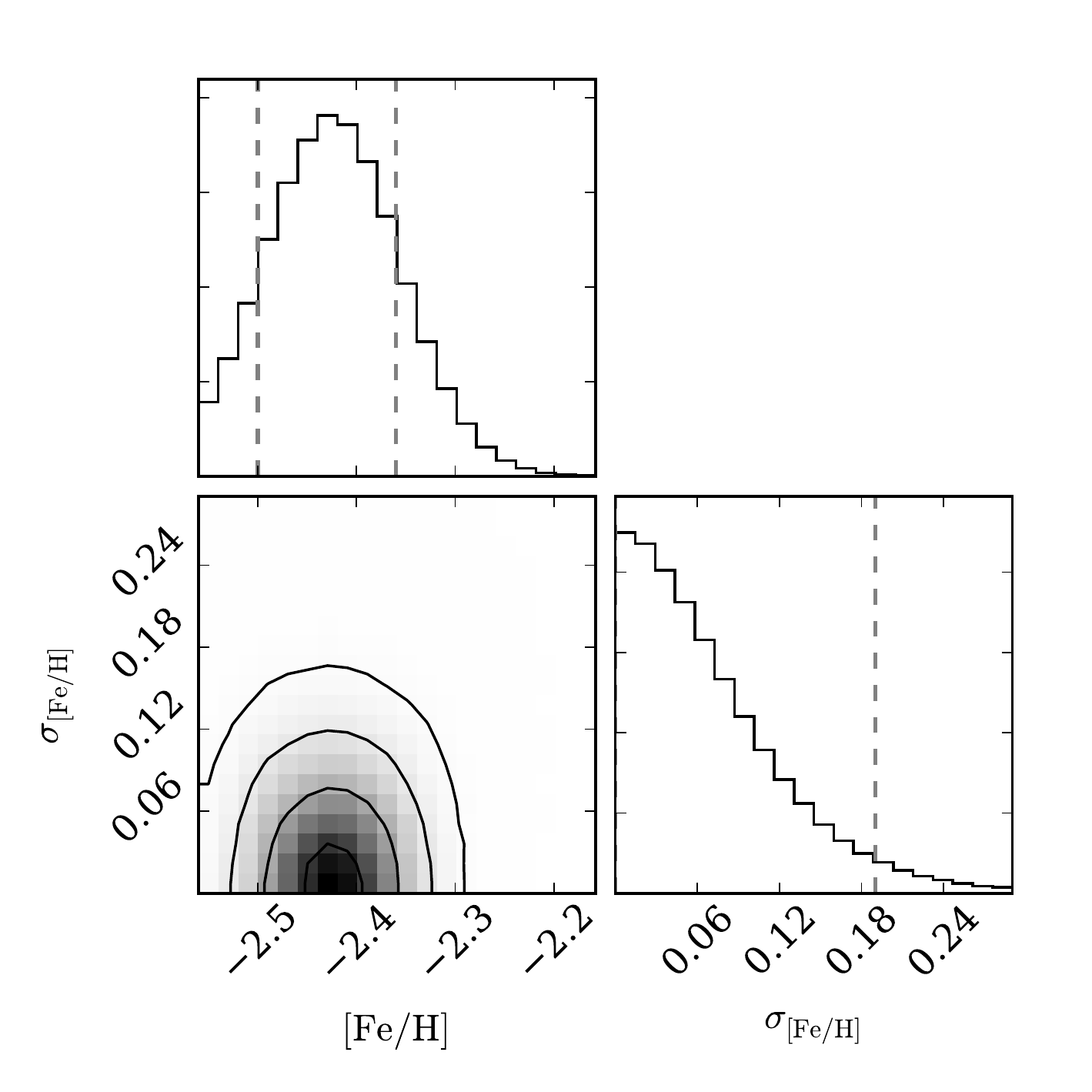}
\caption{Posterior probability distribution from a maximum likelihood fit for the systemic velocity and velocity dispersion (left set of panels) and the mean metallicity and metallicity dispersion (right set of panels) of Tuc~III.  In the upper left panels the $68\%$ confidence intervals on the mean velocity and metallicity are indicated by the dashed gray lines.  In the lower right panels the $95.5\%$ upper limits on the velocity and metallicity dispersion are indicated by the dashed gray lines.  We do not significantly resolve either the velocity dispersion or the metallicity dispersion of Tuc~III.}
\label{vmcmc}
\end{figure*}

In general, it is possible that binary stars can affect velocity dispersion
measurements for kinematically cold systems.
Previous studies indicate that binary stars generally do not substantially inflate the observed
velocity dispersions of ultra-faint dwarfs \citep{minor10,simon11}, but the
smaller the dispersion of an object the larger the impact of the
binaries could be \citep{mc10}.  Approximate radial velocity
amplitudes are only known for a handful of ultra-faint dwarf RGB
binaries, but typical orbital velocities and periods appear to be
$\sim30$~\kms\ and a few months, respectively
\citep{koposov11,koch14,ji16a}.  We have obtained multiple velocity measurements spaced $\sim1$~yr apart for eleven of the Tuc~III member stars, including eight of the ten RGB stars, which have the smallest velocity uncertainties and are therefore the most important in determining the velocity dispersion.  For eight of the eleven stars with repeat measurements the velocities agree within the uncertainties, and for the other three the velocity difference is between 1 and 1.5$\sigma$.  These results are exactly in accordance with expectations from Gaussian statistics if all of the stars have velocities that are constant with time.  We conclude that there is no evidence that our Tuc~III member sample contains any binaries with short enough periods or large enough amplitudes to affect our constraint on the velocity dispersion.

We use the upper limit on the velocity dispersion of Tuc~III along with its half-light radius to 
constrain the mass contained within its half-light radius according
to the formula derived by \citet{wolf10}.  Ignoring the uncertainty on the radius (which is only 14\%; \citealt{dw15b}), we find 90\%, 95.5\%, and 99.7\%\ upper limits of $M_{1/2} \le 
6.0 \times 10^{4}$~M$_{\odot}$, $M_{1/2} \le 
9.0 \times 10^{4}$~M$_{\odot}$, and $M_{1/2} \le 
2.2 \times 10^{5}$~M$_{\odot}$, respectively.
Given a luminosity within the same radius of 390~L$_{\odot}$, Tuc~III could have a
mass-to-light ratio as large as 240~\msun/\lsun at $2\sigma$ (or 580~\msun/\lsun at $3\sigma$).  The dynamical mass-to-light ratio would be larger than 10 as long as $M_{1/2} >
3900$~M$_{\odot}$.  Using the \citeauthor{wolf10} relation, $M_{1/2}/L_{1/2} > 10$ implies $\sigma > 0.3$~\kms.  Despite its very small velocity dispersion, Tuc~III can easily be dynamically dominated by dark matter.  However, firmly ruling out the alternative that Tuc~III consists solely of baryons will only be possible with extremely accurate high-resolution spectroscopy of a significant sample of Tuc~III stars.

Even though we are not able to directly measure the dynamical mass of Tuc~III, we can estimate its mass via an indirect argument.  Tuc~III is currently located 23~kpc from the Galactic Center, and for a Galactic rotation velocity of 220~\kms\ \citep[e.g.,][]{bovy12} the Milky Way mass enclosed out to that radius is $2.6 \times 10^{11}$~\msun.  The Jacobi (tidal) radius of Tuc~III is therefore $r_{J} = 23000~(M_{{\rm Tuc III}}/2.6 \times 10^{11}~\mbox{M}_{\odot})^{1/3}$~pc \citep{bt08}.  The 90\%, 95.5\%, and 99.7\%\ upper limits on the Jacobi radius are then 142, 162, and 219~pc, respectively.  The tidal tails of Tuc~III are visible by a radius of 0.3\degr\ (131~pc), and may extend inward as far as 0.2\degr\ (87~pc), consistent with these upper limits.  On the other hand, if Tuc~III had a mass-to-light ratio of 1~\msun/\lsun\ (M$_{1/2} = 390$~\msun) the Jacobi radius would be just 27~pc, which is not consistent with the observational result that the stars exhibit no detectable velocity gradient or dispersion out to a radius of $\sim 90$~pc.  If we demand based on the velocity data that $r_{J} > 90$~pc, the implied lower limit on the mass of Tuc~III is $1.6 \times 10^{4}$~\msun\ (giving $M/L > 20$~\msun/\lsun).

N-body simulations suggest that the dynamical mass-to-light ratio
of a dwarf galaxy is only modestly affected by tidal stripping
until more than 90\%\ of the original mass of the system has been
lost \citep*{penarrubia08}.  After that point, stripping actually 
increases the mass-to-light ratio because the central dark matter 
cusp is more tightly bound than the stars.  The tidal tails around
Tuc~III have a similar luminosity to the bound core, suggesting 
that at least $\sim50$\%\ of the stars initially in the system have been stripped.
According to the \citeauthor{penarrubia08} models, the change in
M/L with that amount of stripping is expected to be quite modest ($\sim30$\%).  Interestingly, if Tuc~III has suffered more 
significant stripping (in which case most of the stars must be in
a more diffuse component of the tails that is below current detection limits), its
unusually low velocity dispersion makes it consistent with a 
$10^{6}$~L$_{\odot}$ progenitor with a 10~\kms\ velocity 
dispersion \citep{penarrubia08}, similar to the classical dwarf spheroidals.  However, its metallicity is lower than would be
expected for such a large dwarf (see \S\ref{sec:metallicity}).  
The larger velocity dispersions of most other ultra-faint dwarfs 
are not consistent with the tidal evolution of more luminous 
systems.

\subsection{Metallicity and Metallicity Spread}
\label{sec:metallicity}

As seen in Figure~\ref{cmd}, ten of the Tuc~III member stars in our sample are on the RGB, with the remainder of the sample on the subgiant branch or the main sequence.
We measure CaT metallicities for the RGB members of Tuc~III,
which range from $\feh = -2.16$ to $\feh = -2.58$.  Using the same MCMC
method as in \S\ref{sec:kinematics}, we calculate a mean metallicity
of $\feh = -2.42^{+0.07}_{-0.08}$, with a dispersion of $\sigma_{\feh} =
0.01^{+0.09}_{-0.01}$.  The metallicity spread in Tuc~III is therefore
unresolved in these data, with an upper limit of 0.19~dex at $95.5\%$ confidence.  The posterior probability distributions from the MCMC fit are displayed in the right panel of Figure~\ref{vmcmc}, and the kinematic and chemical  properties of Tuc~III are summarized in Table~\ref{tuc3_table}.

\input{tuc3_table.tex}

The metallicity range of the confirmed Tuc~III members is smaller than
that of other dwarf galaxies, especially those with similar luminosities
\citep{kirby13b}.  Other systems with $L < 10^{4}$~L$_{\odot}$ such as
Segue~2, Coma~Berenices, and Ursa~Major~II have metallicity spreads of
$\sim0.4-0.6$~dex, although some of those dwarfs also have larger samples of
stars from which the spread can be measured.  At 95.5\%\ confidence we 
cannot rule out a spread of $\sim0.2$~dex in Tuc~III, so given the currently
available data we conclude that the apparently narrow metallicity 
range of Tuc~III stars is most likely a coincidence resulting from the
small sample size.  If this result persists with a larger
member sample or with more accurate measurements then other hypotheses may need to be considered.

\subsection{The Nature of Tucana III}

\citet{ws12} suggested that a stellar system could be classified as a
galaxy if it has either dynamical evidence for the presence of dark
matter or chemical evidence for the retention of supernova ejecta.  This chemical evidence would consist of an internal spread in the abundance of iron or other heavy elements formed only in explosive events.  Unfortunately, since we are only able to place upper limits on the dynamical mass and metallicity spread, we cannot classify Tuc~III with these standard criteria.

The stellar kinematics of Tuc~III (\S\ref{sec:kinematics}) are consistent either with a dark matter-dominated system if $\sigma \gtrsim 0.3$~\kms\ or a baryon-dominated one if $\sigma \lesssim 0.3$~\kms.  Tuc~III therefore cannot be robustly classified based on
its kinematics.  Even if the velocity dispersion is near the upper limits we derive, it is worth noting that recent numerical simulations of disrupting clusters suggest that they can briefly reach velocity dispersions of $\sim1.5$~\kms\ and half-light radii above 20~pc just before complete disruption occurs (Contenta et al. 2016, in prep.).

The metallicity spread within Tuc~III (\S\ref{sec:metallicity}) is also consistent with either a dark matter-dominated dwarf galaxy or a baryon-dominated cluster.  Its metallicity spread appears to be
smaller than those of other low-luminosity dwarfs, but with only 
CaT-based metallicities and our limited sample of stars we cannot
rule out a substantial spread in \feh.  

\begin{figure*}[th!]
\center
\includegraphics[width=0.75\linewidth]{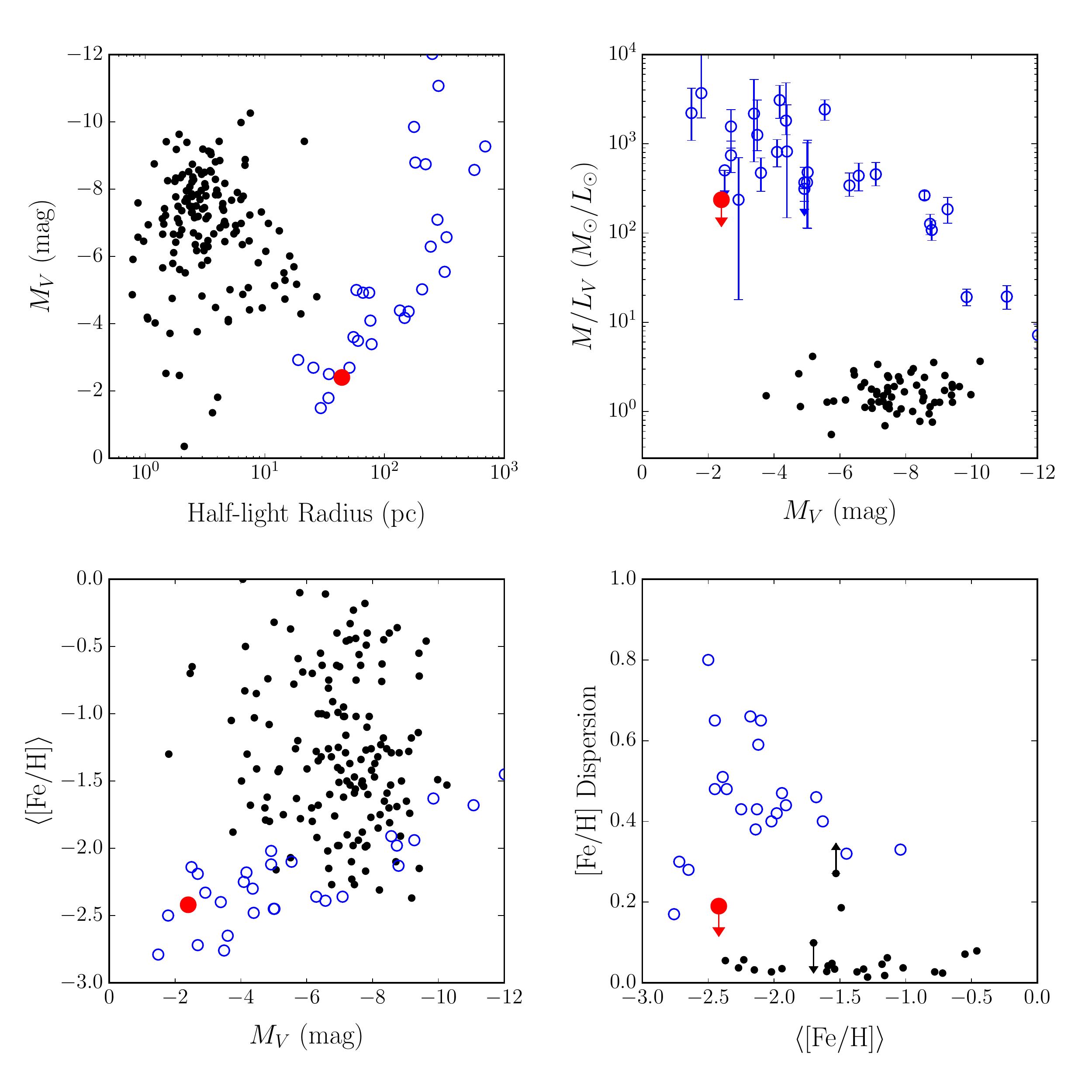}
\caption{Position of Tuc~III relative to dwarf galaxy scaling relations. \emph{(upper left)} Distribution of Milky Way globular clusters (small filled black points) and spectroscopically-confirmed Milky Way satellite galaxies (open blue circles) in the plane of elliptical half-light radius and absolute magnitude.  Tuc~III is plotted as a filled red circle.  \emph{(upper right)} Dynamical mass-to-light ratios for dwarf galaxies and globular clusters as a function of absolute magnitude.  Masses shown are calculated as the mass within the elliptical half-light radius using the formula from \citet{wolf10}.  Symbols are as in the upper left panel except that Tuc~III and Segue~2 are plotted as upper limits.  \emph{(lower left)} The luminosity-metallicity relation for dwarf galaxies, and lack thereof for globular clusters.  Symbols are the same as in the upper left panel.  \emph{(lower right)} Metallicity dispersion as a function of metallicity for dwarf galaxies and globular clusters.  Symbols are the same as in the upper left panel.
The globular cluster data shown in this figure are taken from the 2010 edition of the \citet{Harris96} catalog, with the exception of the metallicity dispersions, which are from \citet{ws12} and references therein.  Half-light radii of the dwarf galaxies are taken from the September 2015 update of the \citet{McConnachie12} compilation, while velocity dispersions are taken from \citet{McConnachie12} for classical dSphs and otherwise from \citet{sg07}, \citet{simon11}, \citet{willman11}, \citet{koposov11,koposov15b}, \citet{kirby13a,kirby15,kirby15b}, \citet{simon15b}, \citet{martin16}, \citet{kim16}, \citet{collins16}, Geha et al.\ (in prep.), and Li et al.\ (in prep.).  Dwarf galaxy metallicities and metallicity dispersions are from \citet{kirby13b}, the velocity dispersion references just listed, and \citet{fsk14}, \citet{brown14}, and \citet{kr14}.}
\label{fig:classify}
\end{figure*}

Given the lack of detection of a velocity or metallicity dispersion, we instead attempt to classify Tuc~III using some of its other properties.  The low mean metallicity of Tuc~III suggests that it is not a globular cluster (Figure~\ref{fig:classify}, left panel).  Among known globular clusters, only the much more luminous clusters M15 
\citep{preston06,sobeck11} and M92 \citep{rs11} have comparable or 
lower metallicities.  Considering just the low luminosity 
($M_{V} > -5$) clusters, the mean metallicity of the objects in the 
2010 edition of the \citet{Harris96} catalog is $\feh = -1.1$ and 
the lowest-metallicity system is Palomar~13, listed at $\feh = -1.88$.  
The only high-resolution abundance measurement for Pal~13 is that of 
\citet{cote02}, who found $\feh = -1.98 \pm 0.31$ for the brightest 
star.  Correcting that metallicity to a modern value for the solar 
iron abundance would increase it to $\feh = -1.81$.  
\citet{bradford11} determined a slightly higher value of $\feh = -1.6$ 
from spectral synthesis modeling of medium-resolution spectra.  The 
metallicity of Tuc~III would make it a significant outlier among 
low-luminosity clusters.  In contrast, the low metallicity of Tuc~III 
fits naturally in the context of ultra-faint dwarf galaxies.  The
metallicity predicted for its luminosity from the universal 
luminosity-metallicity relation of \citet{kirby13b} is $\feh = -2.58$.  
Tuc~III is about $1\sigma$ to the high side of this relation, 
consistent with the idea that what we are observing today is the 
remnant of a system that used to be more luminous.  Placing Tuc~III exactly on the luminosity-metallicity relation would result in a luminosity of $2600$~\lsun, implying that $\sim70$\%\ of Tuc~III's stars may have been removed by tidal stripping.

Another diagnostic that can be used to distinguish dwarf galaxies
from globular clusters is size: it has been known for many years
that globular clusters have smaller radii than dwarfs at the
same luminosity \citep[e.g.,][]{belokurov07}.  A half-light
radius of $44 \pm 6$~pc and corresponding surface brightness of 
28.7~mag~arcsec$^{-2}$ places Tuc~III firmly within the dwarf 
galaxy locus, more extended than the similarly faint dwarfs 
Segue~1, Willman~1, Segue~2, and Triangulum~II, as well 
as all low-luminosity globular clusters (Fig.~\ref{fig:classify}, middle panel).  Unless the size of 
Tuc~III has been significantly inflated as a result of tidal 
stripping, which is contrary to expectations from numerical
simulations \citep{penarrubia08}, its radius alone strongly
suggests that it is a galaxy.

The shape of Tuc~III is also more consistent with the population of faint dwarf galaxies than globular clusters.  While \citet{dw15b} were not able to constrain its shape photometrically because of the extension of the tidal tails to relatively small radii, our spectroscopic member sample strongly suggests that Tuc~III is highly elongated (see Figure~\ref{cmd}b).  One might assume that this shape is a result of our lack of spectroscopic coverage in the southwest portion of Tuc~III (Fig.~\ref{cmd}b), but in fact the reason we did not observe that area is that we found no photometric RGB candidates there.
As first noted by \citet{martin08}, ultra-faint dwarfs have systematically larger ellipticities than both brighter dwarfs and globular clusters.  It is widely assumed that very elongated shapes are a telltale sign of tidal disruption, but both simulations \citep{munoz08} and observations (e.g., Pal~5; \citealt{martin08}) argue to the contrary.  Nevertheless, some disrupting systems are quite elliptical (such as Sagittarius), and it is possible that the shape of Tuc~III is more closely related to its dynamical state than to its nature.  

While the arguments presented here are not definitive, the 
combination of its low mean metallicity and structure and the difficulty of reconciling the lack of a velocity dispersion or gradient with tidal disruption inside the observed region indicates that Tuc~III is most likely a dwarf galaxy rather than a globular cluster.  This classification can be confirmed by much more accurate ($\sim0.1$~\kms) radial velocity measurements or by additional and/or more accurate metallicities demonstrating the existence of a spread of 
iron abundances.

If Tuc~III is not a cluster, a possible explanation for its apparently very small metallicity spread is the tidal stripping 
the system has suffered.  Several known dwarfs have
been found to contain multiple stellar populations, with the
metal-rich stars being centrally concentrated and the metal-poor
ones more spatially extended 
\citep[e.g.,][]{tolstoy04,faria07,battaglia11,deboer11,santana16}.  
If the same were true for Tuc~III then the metal-poor stars would 
be more likely to be stripped, increasing the mean metallicity and 
decreasing the metallicity dispersion of the remnant core of the 
system.  Thus, while we would generally expect Tuc~III to have a significant metallicity dispersion if it is a dwarf galaxy, the observed small dispersion is not necessarily
inconsistent with a galactic classification if tidal stripping has altered the original metallicity distribution.
This scenario can be tested by determining the 
metallicity of stars in the tidal tails.

Under the assumption that Tuc~III is indeed a galaxy, it is one
of the closest known dwarfs to the Sun.  Until the recent flood of
discoveries, the nearest dwarfs were Segue~1, at $d = 23 \pm 2$~kpc 
\citep{belokurov07}, and the Sagittarius dSph, at
$d = 26.7 \pm 1.3$~kpc \citep{hamanowicz16}.  The maximum-likelihood
fit to the DES photometry by \citet{dw15b} places Tuc~III in between
the two, at $d = 25 \pm 2$~kpc, but a direct comparison of the
Segue~1 and Tuc~III CMDs suggests that the distances are
indistiguishable without deeper photometry.  Draco~II, whose classification is unclear \citep{martin16}, is also at a similar distance from the Sun ($d = 24$~kpc; \citealt{laevens15b}).  Of these objects, only Sagittarius is closer to the Galactic Center than Tuc~III, so it is perhaps not surprising that those are the two satellites with the clearest signs of tidal disruption.

\subsection{Constraints On the Orbit of Tuc~III}

The Galactocentric velocity that we measure for Tuc~III is 
$v_{GSR} = -195.2 \pm 0.4 \pm 2.0$~\kms.  The large absolute value of the velocity 
suggests that we are observing Tuc~III relatively close to the pericenter 
of its orbit around the Milky Way.  Possible orbital configurations that
would result in a negative radial velocity include (1) the pericenter is 
on the near side of the Galaxy relative to the Sun, with Tuc~III currently
approaching pericenter, (2) the pericenter is on the far side of the Galaxy
and Tuc~III is approaching pericenter, or (3) the pericenter is on the far 
side of the Galaxy and Tuc~III has just passed pericenter.  However, 
given the position of Tuc~III ($\ell = 315.38\degr, b = -56.19\degr$ in 
Galactic coordinates), this last configuration requires a substantial tangential 
velocity for Tuc~III, which would result in a very high space velocity.  
Without a proper motion for the system we cannot currently distinguish 
between these geometries, although radial velocity measurements in the 
tidal tails could provide additional constraints.  




The observed velocity of Tuc~III differs from the predictions of 
\citet{jethwa16} assuming that Tuc~III is a satellite of the Magellanic 
Clouds.  The heliocentric velocity of Tuc~III differs by 150~\kms\ from 
that predicted for an LMC satellite and by 380~\kms from the prediction 
for an SMC satellite.  However, the large uncertainties on the predictions 
of \citeauthor{jethwa16} mean that the observed heliocentric velocity of 
Tuc~III lies just outside the 68\% confidence interval for being a 
satellite of the LMC.  We therefore find that Tuc~III is unlikely to be a 
satellite of the SMC or LMC, but that we cannot rule out these 
scenarios with high confidence.  Consistent with this result, \citet{sales16}
find that the position and distance of Tuc~III give it a low probability
of being associated with the LMC.


\subsection{J-Factor}
We derive an astrophysical \Jfactor for dark matter annihilation  by modeling the velocity distribution using the spherical Jeans equation \citep[\eg,][]{Strigari:2007at,Essig:2009jx,Charbonnier:2011ft,Martinez:2013ioa,Geringer-Sameth:2014yza, simon15b}.
Here, we model the dark matter halo as a generalized
Navarro-Frenk-White (NFW) profile \citep{nfw96}. 
We use flat, `uninformative' priors on the dark matter halo parameters \citep[see][]{Essig:2009jx} and assume a constant stellar velocity anisotropy and a tidal radius of 100~pc.  
From this procedure, we find an upper limit on the \Jfactor for Tuc~III of $\log_{10}(J) < 17.8$ at 90\%\ confidence within an angular cone of radius $0.2\degr$.  However, if we adopt
the lower limit on the mass of Tuc~III from the tidal radius argument in \S\ref{sec:kinematics}, \Jfactors as large as $\log_{10}(J) = 19.4$ are allowed.
Because the tidal tails of Tuc~III indicate that its dark matter halo is truncated at a radius of less than 0.5\degr\ (see \S\ref{sec:kinematics}) we do not calculate \Jfactors for larger angular extents.

The upper limit on the \Jfactor of Tuc~III is more than an order of magnitude below the value predicted from a simple distance scaling based on the \Jfactors of known dwarfs \citep[e.g.,][]{dw15a}.
This low \Jfactor is a direct result of the small velocity dispersion and mass of Tuc~III, which are both lower than would be expected from simple scaling relations.
Thus, in spite of its proximity, Tuc~III does not appear to be a particularly promising target for indirect searches for dark matter annihilation.

\section{SUMMARY AND CONCLUSIONS}
\label{conclusions}

We obtained medium-resolution Magellan/IMACS spectroscopy of stars in
the recently-discovered Milky Way satellite Tucana~III.  By extending 
techniques developed for Keck/DEIMOS observations of similar systems,
we show that it is possible to measure velocities with an accuracy of
$\sim1$~\kms\ with the IMACS 1200~$\ell$/mm grating.  Based on extensive
radial velocities measurements in the vicinity of Tuc~III we identify 26 
stars as Tuc~III members, including ten stars on the red giant branch.  The brightest several member
stars are well within range for high-resolution spectroscopy to
determine chemical abundances, and the most luminous star in Tuc~III
($V=15.7$) is currently the brightest known star in an ultra-faint
dwarf galaxy.

We find a mean velocity for Tuc~III of $v_{hel} = 
-102.3 \pm 0.4~\mbox{(stat.)} \pm 2.0~\mbox{(sys.)}$~\kms\ ($v_{GSR} = -195.2 \pm 0.4 \pm 2.0$~\kms) 
and a velocity dispersion of $0.1^{+0.7}_{-0.1}$~\kms.  The 95.5\%\ and 99.7\%\ confidence upper limits on the velocity dispersion are 1.5~\kms\ and 2.3~\kms, respectively.  The mass within the half-light radius of Tuc~III is therefore less than $9 \times 10^{4}$~M$_{\odot}$ at 95.5\%\ confidence, corresponding to an upper limit on the dynamical mass-to-light ratio of 240~M$_{\odot}$/L$_{\odot}$.  Among ultra-faint stellar systems for which kinematic measurements are available, only Segue~2 ($\sigma < 2.2$~\kms at 90\%\ confidence; \citealt{kirby13a}) has a comparably small velocity dispersion.

Tuc~III has a mean metallicity of $\feh = -2.42^{+0.07}_{-0.08}$, with an
unresolved spread of $\sigma_{\feh} < 0.19$~dex at 95.5\% confidence.  We conclude that the low metallicity, large radius, elongated shape, and lack of detectable kinematic disturbances out to a radius of $\sim90$~pc are most consistent with a dwarf galaxy classification for Tuc~III, with a small residual dark matter component.  
Tuc~III is therefore one of the closest dwarfs to the Sun.
The placement of Tuc~III relative to the luminosity-metallicity relation and its small metallicity dispersion may be the result of extensive tidal stripping.  If interpreted instead as a globular cluster, it would be the most extended cluster known, with a metallicity $\gtrsim0.5$~dex lower than any similar-luminosity cluster.  Further metallicity measurements to determine the iron abundance spread or detailed chemical abundance patterns for the brightest member stars can definitively confirm our classification.

The large negative velocity of Tuc~III suggests that it is currently
close to the pericenter of its orbit around the Milky Way.  This
velocity is quite different from that expected if Tuc~III originated
as a Magellanic satellite that has been accreted by the Milky Way,
so we conclude that Tuc~III probably did not form in the Magellanic
group.  Measurements of the proper motion of Tuc~III and the 
velocity gradient in its tidal tails will constrain its orbit more
tightly and firmly determine its origin.

\acknowledgements{This publication is based upon work supported by the
  National Science Foundation under grant AST-1108811.  We thank Dan
  Kelson for many helpful conversations regarding IMACS data
  reduction.  E.B. acknowledges financial support from the European Research Council
(ERC-StG-335936).  This research has made use of NASA's Astrophysics Data
  System Bibliographic Services. Contour plots were generated using \code{corner.py} \citep{corner}.

  Funding for the DES Projects has been provided by the
  U.S. Department of Energy, the U.S. National Science Foundation, the
  Ministry of Science and Education of Spain, the Science and
  Technology Facilities Council of the United Kingdom, the Higher
  Education Funding Council for England, the National Center for
  Supercomputing Applications at the University of Illinois at
  Urbana-Champaign, the Kavli Institute of Cosmological Physics at the
  University of Chicago, Financiadora de Estudos e Projetos, Funda{\c
    c}{\~a}o Carlos Chagas Filho de Amparo {\`a} Pesquisa do Estado do
  Rio de Janeiro, Conselho Nacional de Desenvolvimento Cient{\'i}fico
  e Tecnol{\'o}gico and the Minist{\'e}rio da Ci{\^e}ncia e
  Tecnologia, the Deutsche Forschungsgemeinschaft and the
  Collaborating Institutions in the Dark Energy Survey.  The DES
  participants from Spanish institutions are partially supported by
  MINECO under grants AYA2012-39559, ESP2013-48274, FPA2013-47986, and
  Centro de Excelencia Severo Ochoa SEV-2012-0234, some of which
  include ERDF funds from the European Union. This material is based
  upon work supported by the National Science Foundation under Grant
  Number (1138766).

  The Collaborating Institutions are Argonne National Laboratory, the
  University of California at Santa Cruz, the University of Cambridge,
  Centro de Investigaciones Energ{\'e}ticas, Medioambientales y
  Tecnol{\'o}gicas-Madrid, the University of Chicago, University College
  London, the DES-Brazil Consortium, the Eidgen{\"o}ssische Technische
  Hochschule (ETH) Z{\"u}rich, Fermi National Accelerator Laboratory,
  the University of Edinburgh, the University of Illinois at
  Urbana-Champaign, the Institut de Ci{\`e}ncies de l'Espai (IEEC/CSIC),
  the Institut de F{\'i}sica d'Altes Energies, Lawrence Berkeley National
  Laboratory, the Ludwig-Maximilians Universit{\"a}t and the
  associated Excellence Cluster Universe, the University of Michigan,
  the National Optical Astronomy Observatory, the University of
  Nottingham, The Ohio State University, the University of
  Pennsylvania, the University of Portsmouth, SLAC National
  Accelerator Laboratory, Stanford University, the University of
  Sussex, and Texas A\&M University.
}

{\it Facilities:} \facility{Magellan:I (IMACS)}


\bibliographystyle{apj}
\bibliography{main}{}

\clearpage 
\input{tuc3_spec_table.tex}

\end{document}

%% file: commands.tex
\newcommand{\eg}{e.g.\xspace}

\mathchardef\mhyphen="2D

\newlength{\dhatheight}

\newcommand{\code}[1]{\texttt{#1}\xspace}

\newcommand{\unit}[1]{\ensuremath{\mathrm{\,#1}}\xspace}

\newcommand{\km}{\unit{km}}

\newcommand{\second}{\unit{s}}

\newcommand{\msun}{\unit{M_\odot}}

\newcommand{\lsun}{\unit{L_\odot}}

\newcommand{\Jfactor}{J-factor\xspace}
\newcommand{\Jfactors}{J-factors\xspace}

\providecommand\physrep{\ref@jnl{Phys.~Rep.}}%
\providecommand\apjs{\ref@jnl{ApJS}}%
\providecommand{\jcap}{\ref@jnl{JCAP}}%

%% file: authors.tex

\def\andname{}

\author{
J.~D.~Simon\altaffilmark{1},
T.~S.~Li\altaffilmark{2,3},
A.~Drlica-Wagner\altaffilmark{2},
K.~Bechtol\altaffilmark{4},
J.~L.~Marshall\altaffilmark{3},
D.~J.~James\altaffilmark{5,6},
M.~Y.~Wang\altaffilmark{3},
L.~Strigari\altaffilmark{3},
E.~Balbinot\altaffilmark{7},
K.~Kuehn\altaffilmark{8},
A.~R.~Walker\altaffilmark{6},
T. M. C.~Abbott\altaffilmark{6},
S.~Allam\altaffilmark{2},
J.~Annis\altaffilmark{2},
A.~Benoit-L{\'e}vy\altaffilmark{9,10,11},
D.~Brooks\altaffilmark{10},
E.~Buckley-Geer\altaffilmark{2},
D.~L.~Burke\altaffilmark{12,13},
A. Carnero Rosell\altaffilmark{14,15},
M.~Carrasco~Kind\altaffilmark{16,17},
J.~Carretero\altaffilmark{18,19},
C.~E.~Cunha\altaffilmark{12},
C.~B.~D'Andrea\altaffilmark{20,21},
L.~N.~da Costa\altaffilmark{14,15},
D.~L.~DePoy\altaffilmark{3},
S.~Desai\altaffilmark{22},
P.~Doel\altaffilmark{10},
E.~Fernandez\altaffilmark{19},
B.~Flaugher\altaffilmark{2},
J.~Frieman\altaffilmark{2,23},
J.~Garc\'ia-Bellido\altaffilmark{24},
E.~Gaztanaga\altaffilmark{18},
D.~A.~Goldstein\altaffilmark{25,26},
D.~Gruen\altaffilmark{12,13},
G.~Gutierrez\altaffilmark{2},
N.~Kuropatkin\altaffilmark{2},
M.~A.~G.~Maia\altaffilmark{14,15},
P.~Martini\altaffilmark{27,28},
F.~Menanteau\altaffilmark{16,17},
C.~J.~Miller\altaffilmark{29,30},
R.~Miquel\altaffilmark{31,19},
E.~Neilsen\altaffilmark{2},
B.~Nord\altaffilmark{2},
R.~Ogando\altaffilmark{14,15},
A.~A.~Plazas\altaffilmark{32},
A.~K.~Romer\altaffilmark{33},
E.~S.~Rykoff\altaffilmark{12,13},
E.~Sanchez\altaffilmark{24},
B.~Santiago\altaffilmark{34,14},
V.~Scarpine\altaffilmark{2},
M.~Schubnell\altaffilmark{30},
I.~Sevilla-Noarbe\altaffilmark{24},
R.~C.~Smith\altaffilmark{6},
F.~Sobreira\altaffilmark{14,35},
E.~Suchyta\altaffilmark{36},
M.~E.~C.~Swanson\altaffilmark{17},
G.~Tarle\altaffilmark{30},
L.~Whiteway\altaffilmark{10},
B.~Yanny\altaffilmark{2}
\\ \vspace{0.2cm} (The DES Collaboration) \\
}
 
\affil{$^{1}$ Carnegie Observatories, 813 Santa Barbara St., Pasadena, CA 91101, USA}
\affil{$^{2}$ Fermi National Accelerator Laboratory, P. O. Box 500, Batavia, IL 60510, USA}
\affil{$^{3}$ George P. and Cynthia Woods Mitchell Institute for Fundamental Physics and Astronomy, and Department of Physics and Astronomy, Texas A\&M University, College Station, TX 77843,  USA}
\affil{$^{4}$ LSST, 933 North Cherry Avenue, Tucson, AZ 85721, USA}
\affil{$^{5}$ Astronomy Department, University of Washington, Box 351580, Seattle, WA 98195, USA}
\affil{$^{6}$ Cerro Tololo Inter-American Observatory, National Optical Astronomy Observatory, Casilla 603, La Serena, Chile}
\affil{$^{7}$ Department of Physics, University of Surrey, Guildford GU2 7XH, UK}
\affil{$^{8}$ Australian Astronomical Observatory, North Ryde, NSW 2113, Australia}
\affil{$^{9}$ CNRS, UMR 7095, Institut d'Astrophysique de Paris, F-75014, Paris, France}
\affil{$^{10}$ Department of Physics \& Astronomy, University College London, Gower Street, London, WC1E 6BT, UK}
\affil{$^{11}$ Sorbonne Universit\'es, UPMC Univ Paris 06, UMR 7095, Institut d'Astrophysique de Paris, F-75014, Paris, France}
\affil{$^{12}$ Kavli Institute for Particle Astrophysics \& Cosmology, P. O. Box 2450, Stanford University, Stanford, CA 94305, USA}
\affil{$^{13}$ SLAC National Accelerator Laboratory, Menlo Park, CA 94025, USA}
\affil{$^{14}$ Laborat\'orio Interinstitucional de e-Astronomia - LIneA, Rua Gal. Jos\'e Cristino 77, Rio de Janeiro, RJ - 20921-400, Brazil}
\affil{$^{15}$ Observat\'orio Nacional, Rua Gal. Jos\'e Cristino 77, Rio de Janeiro, RJ - 20921-400, Brazil}
\affil{$^{16}$ Department of Astronomy, University of Illinois, 1002 W. Green Street, Urbana, IL 61801, USA}
\affil{$^{17}$ National Center for Supercomputing Applications, 1205 West Clark St., Urbana, IL 61801, USA}
\affil{$^{18}$ Institut de Ci\`encies de l'Espai, IEEC-CSIC, Campus UAB, Carrer de Can Magrans, s/n,  08193 Bellaterra, Barcelona, Spain}
\affil{$^{19}$ Institut de F\'{\i}sica d'Altes Energies (IFAE), The Barcelona Institute of Science and Technology, Campus UAB, 08193 Bellaterra (Barcelona) Spain}
\affil{$^{20}$ Institute of Cosmology \& Gravitation, University of Portsmouth, Portsmouth, PO1 3FX, UK}
\affil{$^{21}$ School of Physics and Astronomy, University of Southampton,  Southampton, SO17 1BJ, UK}
\affil{$^{22}$ Department of Physics, IIT Hyderabad, Kandi, Telangana 502285, India}
\affil{$^{23}$ Kavli Institute for Cosmological Physics, University of Chicago, Chicago, IL 60637, USA}
\affil{$^{24}$ Centro de Investigaciones Energ\'eticas, Medioambientales y Tecnol\'ogicas (CIEMAT), Madrid, Spain}
\affil{$^{25}$ Department of Astronomy, University of California, Berkeley,  501 Campbell Hall, Berkeley, CA 94720, USA}
\affil{$^{26}$ Lawrence Berkeley National Laboratory, 1 Cyclotron Road, Berkeley, CA 94720, USA}
\affil{$^{27}$ Center for Cosmology and Astro-Particle Physics, The Ohio State University, Columbus, OH 43210, USA}
\affil{$^{28}$ Department of Astronomy, The Ohio State University, Columbus, OH 43210, USA}
\affil{$^{29}$ Department of Astronomy, University of Michigan, Ann Arbor, MI 48109, USA}
\affil{$^{30}$ Department of Physics, University of Michigan, Ann Arbor, MI 48109, USA}
\affil{$^{31}$ Instituci\'o Catalana de Recerca i Estudis Avan\c{c}ats, E-08010 Barcelona, Spain}
\affil{$^{32}$ Jet Propulsion Laboratory, California Institute of Technology, 4800 Oak Grove Dr., Pasadena, CA 91109, USA}
\affil{$^{33}$ Department of Physics and Astronomy, Pevensey Building, University of Sussex, Brighton, BN1 9QH, UK}
\affil{$^{34}$ Instituto de F\'\i sica, UFRGS, Caixa Postal 15051, Porto Alegre, RS - 91501-970, Brazil}
\affil{$^{35}$ Universidade Federal do ABC, Centro de Ci\^encias Naturais e Humanas, Av. dos Estados, 5001, Santo Andr\'e, SP, Brazil, 09210-580}
\affil{$^{36}$ Computer Science and Mathematics Division, Oak Ridge National Laboratory, Oak Ridge, TN 37831}

%% file: obstable.tex
\begin{deluxetable*}{lcccrccc}
\tablecaption{Observations}
\tablewidth{0pt}
\tablehead{
\colhead{Mask} &
\colhead{$\alpha$ (J2000)} &
\colhead{$\delta$ (J2000)} &
\colhead{Slit PA} &
\colhead{$t_{\rm exp}$} &
\colhead{MJD of} &
\colhead{\# of slits} &
\colhead{\% useful} \\
\colhead{name}&
\colhead{(h$\,$ $\,$m$\,$ $\,$s)} &
\colhead{($^\circ\,$ $\,'\,$ $\,''$)} &
\colhead{(deg)} &
\colhead{(sec)} &
\colhead{observation\tablenotemark{a}} &
\colhead{} &
\colhead{spectra}
}
\startdata
Mask 1     & 23 56 49.50 & $-59$ 34 40.0  & 292.0 & 23400 & 57220.8 & 63 & \phn54\%\\
Mask 2     & 23 57 15.90 & $-59$ 34 34.0  & 172.0 & 3600  & 57223.3 & 58 & \phn17\%\\
           &             &                &       & 2400  & 57312.2 &    & \phn36\%\\
Longslit   & 23 55 23.60 & $-59$ 30 34.8  & 128.6 & 2700  & 57223.3 & \phn2 & 100\%\\
Mask 3     & 23 56 48.00 & $-59$ 35 10.0  & 297.0 & 37200 & 57630.9 & 64 & \phn70\%\\
Mask 4     & 23 58 20.00 & $-59$ 38 00.0  & 172.0 & 3600  & 57632.1 & 30 & \phn63\%\\
Mask 5     & 23 55 13.00 & $-59$ 31 45.0  & 221.0 & 5400  & 57632.8 & 34 & \phn62\%\\
\enddata
\tablenotetext{a}{For observations made over multiple nights, the date listed here is 
the weighted mean observation date, which may occur during daylight hours.  }
\label{obstable}
\end{deluxetable*}

%% file: tuc3_table.tex
\begin{deluxetable}{llr}
\tablecaption{Summary of Properties of Tucana\,III}
\tablenum{3}
\tablewidth{0pt}
\tablehead{
\colhead{Row} & \colhead{Quantity} & \colhead{Value}
}
\startdata
(1) & RA (J2000)                           & 23:56:36 \\
(2) & Dec (J2000)                          & $-59$:36:00 \\
(3) & Distance (kpc)                       & $25 \pm 2$  \\
(4) & $M_{V,0}$                             & $-2.4 \pm 0.4$ \\
(5) & $L_{V,0}$ (L$_{\odot}$)               & $780^{+350}_{-240}$ \\
(6) & $r_{\rm 1/2}$ (pc)                    & $44 \pm 6$  \\ 
[+1ex] \tableline \\ [-1ex]
(7)  & $V_{\rm hel}$ (\kms)                  & $-102.3 \pm 0.4 \pm 2.0$ \\
(8)  & $V_{\rm GSR}$ (\kms)                  & $-195.2 \pm 0.4 \pm 2.0$ \\
(9)  & $\sigma$ (\kms)\tablenotemark{a}      & $<1.5$ \\
(10)  & Mass (M$_{\odot}$)\tablenotemark{a}  & $<8 \times 10^{4}$  \\
(11)  & M/L$_{V}$ (M$_{\odot}$/L$_{\odot}$)\tablenotemark{a}  & $<240$ \\
(12)  & Mean [Fe/H]                        & $-2.42^{+0.07}_{-0.08}$ \\
(13)  & Metallicity dispersion (dex)\tablenotemark{a}       & $<0.19$ \\
(14)  & $\log_{10}{J(0.2\degr)}$ (GeV$^{2}$~cm$^{-5}$) & $<17.8$ \\

\enddata

\tablecomments{Rows (1)-(6) are taken from the DES photometric
  analysis of \citet{dw15b}.  Values in rows (7)-(14) are derived
  in this paper.}
  \tablenotetext{a}{Upper limits listed here are at 95.5\%\ confidence.  See the text for values at other confidence levels.}
\label{tuc3_table}
\end{deluxetable}

%% file: tuc3_spec_table.tex
\LongTables

\tabletypesize{\scriptsize}
\begin{deluxetable*}{c c c c c c c r c c c}
\tablecaption{Velocity and metallicity measurements for Tucana III.\label{tab:tuc3_spec}}
\tablenum{2}

\tablehead{ID & MJD & RA & DEC & $g$\tablenotemark{a} & $r$\tablenotemark{a} & S/N & \multicolumn{1}{c}{$v$} & ${\rm EW}$ & ${\rm [Fe/H]}$ & MEM \\ 
 &  & (deg) & (deg) & (mag) & (mag) &  & \multicolumn{1}{c}{(\kms)} & (\AA) &  &  }
\startdata

DES\,J235421.50$-$593432.2 & 57632.8 & 358.58958 & -59.57561 & 19.32 & 19.05 &  6.5 & $10.66 \pm 4.08$ & $1.53 \pm 0.43$ & -- &    0\\
DES\,J235440.00$-$593807.2 & 57632.8 & 358.66667 & -59.63533 & 19.96 & 19.57 &  4.4 & $297.99 \pm 7.77$ & -- & -- &    0\\
DES\,J235443.22$-$593753.1 & 57632.8 & 358.68007 & -59.63142 & 19.71 & 19.36 &  5.3 & $17.03 \pm 2.74$ & $4.88 \pm 0.67$ & -- &    0\\
DES\,J235453.12$-$593505.3 & 57632.8 & 358.72131 & -59.58481 & 18.47 & 17.93 & 21.3 & $10.54 \pm 1.24$ & -- & -- &    0\\
DES\,J235456.88$-$592811.2 & 57632.8 & 358.73701 & -59.46979 & 17.79 & 17.19 & 48.9 & $77.36 \pm 1.03$ & $5.85 \pm 0.34$ & -- &    0\\
DES\,J235457.61$-$593733.9 & 57632.8 & 358.74004 & -59.62609 & 16.57 & 15.94 & 59.5 & $48.93 \pm 1.04$ & $6.13 \pm 0.34$ & -- &    0\\
DES\,J235459.50$-$593510.4 & 57632.8 & 358.74790 & -59.58622 & 21.52 & 20.60 &  3.2 & $82.51 \pm 7.87$ & $7.57 \pm 1.29$ & -- &    0\\
DES\,J235500.62$-$593035.5 & 57632.8 & 358.75257 & -59.50986 & 17.57 & 17.28 & 39.9 & $-1.04 \pm 1.09$ & $3.18 \pm 0.43$ & -- &    0\\
DES\,J235500.75$-$593157.8 & 57632.8 & 358.75314 & -59.53272 & 20.09 & 19.72 &  6.8 & $-103.30 \pm 1.91$ & $1.94 \pm 0.47$ & -- &    1\\
DES\,J235500.98$-$593459.0 & 57632.8 & 358.75409 & -59.58307 & 19.26 & 18.82 & 11.4 & $-40.92 \pm 1.52$ & $2.79 \pm 0.50$ & -- &    0\\
DES\,J235501.69$-$593126.3 & 57632.8 & 358.75706 & -59.52398 & 18.98 & 18.62 & 15.8 & $ 4.58 \pm 1.30$ & $3.31 \pm 0.45$ & -- &    0\\
DES\,J235514.47$-$592929.2 & 57632.8 & 358.81029 & -59.49145 & 19.91 & 19.47 & 10.8 & $-217.68 \pm 1.71$ & $5.53 \pm 0.61$ & -- &    0\\
DES\,J235514.96$-$593037.6 & 57632.8 & 358.81232 & -59.51044 & 21.63 & 21.00 &  3.0 & $-41.99 \pm 11.23$ & -- & -- &    0\\
DES\,J235515.82$-$592848.7 & 57632.8 & 358.81591 & -59.48019 & 20.07 & 19.68 &  9.7 & $-102.39 \pm 2.76$ & -- & -- &    1\\
DES\,J235515.86$-$592703.6 & 57632.8 & 358.81608 & -59.45100 & 20.98 & 20.21 &  6.6 & $90.16 \pm 2.96$ & $4.97 \pm 1.11$ & -- &    0\\
DES\,J235523.86$-$592926.5 & 57632.8 & 358.84940 & -59.49070 & 19.21 & 18.86 & 16.4 & $65.15 \pm 1.31$ & $2.05 \pm 0.41$ & -- &    0\\
DES\,J235526.38$-$592747.9 & 57632.8 & 358.85993 & -59.46330 & 19.78 & 19.40 & 11.3 & $ 2.62 \pm 1.55$ & $3.38 \pm 0.55$ & -- &    0\\
DES\,J235532.66$-$593114.9 & 57223.3 & 358.88609 & -59.52081 & 16.09 & 15.36 & 34.7 & $-102.32 \pm 1.23$ & $3.83 \pm 0.35$ & $-2.24 \pm 0.15$ &    1\\
... & 57632.8 & ... & ... & ... & ... & 130.2 & $-103.26 \pm 1.00$ & $3.57 \pm 0.32$ & $-2.35 \pm 0.14$ & ...\\
DES\,J235534.70$-$592520.3 & 57632.8 & 358.89458 & -59.42232 & 19.62 & 19.01 & 15.7 & $73.79 \pm 2.06$ & $5.24 \pm 0.54$ & -- &    0\\
DES\,J235536.68$-$592552.4 & 57632.8 & 358.90281 & -59.43122 & 21.90 & 21.19 &  3.0 & $74.31 \pm 9.96$ & -- & -- &    0\\
DES\,J235542.74$-$592710.6 & 57632.8 & 358.92807 & -59.45293 & 18.57 & 18.27 & 22.6 & $51.17 \pm 1.20$ & $2.97 \pm 0.38$ & -- &    0\\
DES\,J235549.90$-$593259.6 & 57220.8 & 358.95790 & -59.54989 & 17.41 & 16.84 & 94.9 & $-102.88 \pm 1.21$ & $2.63 \pm 0.33$ & $-2.46 \pm 0.17$ &    1\\
... & 57223.3 & ... & ... & ... & ... &  7.2 & $-101.81 \pm 2.90$ & $2.29 \pm 0.66$ & $-2.63 \pm 0.34$ & ...\\
... & 57630.9 & ... & ... & ... & ... & 142.6 & $-101.92 \pm 1.00$ & $2.66 \pm 0.32$ & $-2.45 \pm 0.16$ & ...\\
DES\,J235555.49$-$593246.3 & 57220.8 & 358.98119 & -59.54619 & 19.00 & 18.18 & 44.2 & $ 5.32 \pm 1.24$ & $5.93 \pm 0.35$ & -- &    0\\
... & 57630.9 & ... & ... & ... & ... & 70.9 & $ 5.67 \pm 1.02$ & $5.90 \pm 0.33$ & -- & ...\\
DES\,J235555.60$-$593156.2 & 57220.8 & 358.98167 & -59.53228 & 17.49 & 17.09 & 71.6 & $83.03 \pm 1.22$ & $4.83 \pm 0.33$ & -- &    0\\
... & 57630.9 & ... & ... & ... & ... & 110.8 & $83.30 \pm 1.01$ & $5.12 \pm 0.32$ & -- & ...\\
DES\,J235602.62$-$593257.8 & 57220.8 & 359.01092 & -59.54939 & 18.94 & 18.47 & 33.6 & $42.05 \pm 1.27$ & $3.83 \pm 0.35$ & -- &    0\\
... & 57630.9 & ... & ... & ... & ... & 51.0 & $40.55 \pm 1.02$ & $4.19 \pm 0.34$ & -- & ...\\
DES\,J235606.52$-$593418.8 & 57220.8 & 359.02717 & -59.57189 & 19.68 & 19.21 & 18.9 & $ 0.89 \pm 1.51$ & $5.46 \pm 0.41$ & -- &    0\\
DES\,J235607.60$-$593022.1 & 57630.9 & 359.03166 & -59.50614 & 21.03 & 20.05 & 19.2 & $158.96 \pm 1.28$ & $4.76 \pm 0.45$ & -- &    0\\
DES\,J235610.63$-$593458.8 & 57630.9 & 359.04430 & -59.58300 & 21.06 & 20.87 &  7.4 & $-103.72 \pm 5.47$ & -- & -- &    1\\
DES\,J235612.07$-$593247.5 & 57630.9 & 359.05029 & -59.54653 & 21.56 & 21.31 &  4.4 & $-102.22 \pm 5.48$ & -- & -- &    1\\
DES\,J235612.82$-$592842.7 & 57630.9 & 359.05344 & -59.47852 & 16.84 & 16.31 & 178.5 & $ 0.11 \pm 1.00$ & $6.24 \pm 0.32$ & -- &    0\\
DES\,J235614.39$-$593313.2 & 57220.8 & 359.05997 & -59.55368 & 20.62 & 20.40 &  5.6 & $-104.75 \pm 2.61$ & -- & -- &    1\\
... & 57630.9 & ... & ... & ... & ... & 11.0 & $-98.54 \pm 3.98$ & -- & -- & ...\\
DES\,J235614.85$-$593022.1 & 57220.8 & 359.06186 & -59.50613 & 19.85 & 19.10 & 23.1 & $43.88 \pm 1.45$ & $4.86 \pm 0.44$ & -- &    0\\
... & 57630.9 & ... & ... & ... & ... & 38.2 & $43.36 \pm 1.06$ & $5.61 \pm 0.36$ & -- & ...\\
DES\,J235615.35$-$593049.8 & 57220.8 & 359.06397 & -59.51384 & 19.44 & 19.05 & 19.7 & $103.70 \pm 1.47$ & -- & -- &    0\\
... & 57630.9 & ... & ... & ... & ... & 33.3 & $111.17 \pm 1.17$ & $4.53 \pm 0.37$ & -- & ...\\
DES\,J235615.42$-$592934.8 & 57220.8 & 359.06426 & -59.49300 & 19.07 & 18.73 & 25.4 & $195.09 \pm 1.33$ & $5.05 \pm 0.40$ & -- &    0\\
... & 57630.9 & ... & ... & ... & ... & 40.3 & $191.65 \pm 1.08$ & -- & -- & ...\\
DES\,J235616.92$-$593045.5 & 57630.9 & 359.07051 & -59.51263 & 20.96 & 20.74 &  7.3 & $63.29 \pm 2.73$ & -- & -- &    0\\
DES\,J235620.18$-$593518.8 & 57630.9 & 359.08409 & -59.58857 & 21.73 & 21.48 &  4.2 & $-106.78 \pm 3.23$ & $1.20 \pm 0.46$ & -- &    1\\
DES\,J235620.75$-$593310.1 & 57220.8 & 359.08645 & -59.55279 & 18.86 & 18.39 & 30.6 & $-102.38 \pm 1.27$ & $2.23 \pm 0.36$ & $-2.36 \pm 0.21$ &    1\\
... & 57630.9 & ... & ... & ... & ... & 57.6 & $-102.82 \pm 1.02$ & $2.10 \pm 0.34$ & $-2.43 \pm 0.20$ & ...\\
DES\,J235624.48$-$593300.4 & 57220.8 & 359.10202 & -59.55010 & 20.46 & 20.25 &  7.0 & $-96.73 \pm 9.11$ & -- & -- &    1\\
... & 57630.9 & ... & ... & ... & ... & 12.0 & $-100.62 \pm 2.04$ & $1.19 \pm 0.53$ & -- & ...\\
DES\,J235628.83$-$593241.3 & 57220.8 & 359.12014 & -59.54481 & 20.45 & 20.23 &  7.2 & $-101.57 \pm 1.93$ & -- & -- &    1\\
DES\,J235632.52$-$593427.1 & 57220.8 & 359.13552 & -59.57420 & 20.17 & 19.83 & 10.0 & $165.35 \pm 2.07$ & $2.77 \pm 0.73$ & -- &    0\\
... & 57630.9 & ... & ... & ... & ... & 17.2 & $160.84 \pm 1.25$ & $3.96 \pm 0.52$ & -- & ...\\
DES\,J235634.87$-$593001.1 & 57630.9 & 359.14530 & -59.50029 & 21.27 & 21.05 &  5.6 & $-104.05 \pm 4.12$ & -- & -- &    1\\
DES\,J235634.88$-$593240.6 & 57630.9 & 359.14531 & -59.54460 & 20.30 & 20.05 & 13.0 & $-100.81 \pm 1.68$ & $2.58 \pm 0.46$ & -- &    1\\
DES\,J235635.09$-$593423.6 & 57220.8 & 359.14619 & -59.57322 & 19.51 & 19.03 & 22.4 & $47.00 \pm 1.38$ & $4.74 \pm 0.52$ & -- &    0\\
... & 57630.9 & ... & ... & ... & ... & 35.6 & $45.33 \pm 1.07$ & $5.30 \pm 0.36$ & -- & ...\\
DES\,J235640.91$-$593301.7 & 57220.8 & 359.17047 & -59.55047 & 20.01 & 19.61 & 14.1 & $234.06 \pm 5.87$ & -- & -- &    0\\
DES\,J235645.16$-$593320.2 & 57630.9 & 359.18818 & -59.55562 & 20.66 & 20.48 & 10.3 & $-105.69 \pm 2.23$ & $1.22 \pm 0.49$ & -- &    1\\
DES\,J235645.76$-$593544.2 & 57630.9 & 359.19068 & -59.59562 & 20.91 & 20.70 &  8.4 & $-127.83 \pm 2.45$ & -- & -- &    0\\
DES\,J235646.83$-$593652.8 & 57630.9 & 359.19511 & -59.61467 & 20.89 & 20.66 &  8.8 & $355.50 \pm 2.58$ & -- & -- &    0\\
DES\,J235650.49$-$593420.9 & 57220.8 & 359.21037 & -59.57247 & 19.94 & 19.51 & 14.1 & $-106.14 \pm 1.55$ & $2.14 \pm 0.41$ & $-2.19 \pm 0.24$ &    1\\
... & 57312.2 & ... & ... & ... & ... &  5.9 & $-103.48 \pm 2.20$ & $2.53 \pm 0.76$ & $-1.97 \pm 0.41$ & ...\\
... & 57630.9 & ... & ... & ... & ... & 27.2 & $-102.96 \pm 1.12$ & $1.66 \pm 0.37$ & $-2.49 \pm 0.27$ & ...\\
DES\,J235652.54$-$593123.1 & 57220.8 & 359.21890 & -59.52308 & 16.61 & 15.96 & 159.5 & $45.87 \pm 1.20$ & $6.04 \pm 0.32$ & -- &    0\\
... & 57223.3 & ... & ... & ... & ... & 44.6 & $46.31 \pm 1.24$ & $6.34 \pm 0.35$ & -- & ...\\
... & 57312.2 & ... & ... & ... & ... & 56.7 & $45.81 \pm 1.22$ & $6.32 \pm 0.34$ & -- & ...\\
DES\,J235655.21$-$593758.0 & 57630.9 & 359.23003 & -59.63278 & 20.36 & 19.46 & 33.8 & $158.24 \pm 1.10$ & $6.14 \pm 0.38$ & -- &    0\\
DES\,J235655.47$-$593707.5 & 57220.8 & 359.23114 & -59.61876 & 19.84 & 19.42 & 16.6 & $-104.40 \pm 1.46$ & $1.63 \pm 0.34$ & $-2.53 \pm 0.25$ &    1\\
... & 57312.2 & ... & ... & ... & ... &  6.3 & $-102.68 \pm 3.12$ & $1.45 \pm 0.41$ & $-2.66 \pm 0.33$ & ...\\
... & 57630.9 & ... & ... & ... & ... & 28.6 & $-101.96 \pm 1.12$ & $1.56 \pm 0.39$ & $-2.58 \pm 0.29$ & ...\\
DES\,J235655.78$-$593641.9 & 57630.9 & 359.23242 & -59.61164 & 21.25 & 21.03 &  6.0 & $-108.37 \pm 3.46$ & $0.94 \pm 0.42$ & -- &    1\\
DES\,J235656.72$-$593518.4 & 57220.8 & 359.23632 & -59.58845 & 19.29 & 18.79 & 29.3 & $41.72 \pm 1.34$ & $5.13 \pm 0.39$ & -- &    0\\
... & 57312.2 & ... & ... & ... & ... &  9.1 & $38.44 \pm 1.85$ & $7.40 \pm 0.64$ & -- & ...\\
DES\,J235656.95$-$594126.5 & 57223.3 & 359.23729 & -59.69070 & 19.15 & 18.67 &  5.8 & $196.01 \pm 2.42$ & $2.37 \pm 0.49$ & -- &    0\\
... & 57312.2 & ... & ... & ... & ... & 10.9 & $195.83 \pm 1.75$ & $2.93 \pm 0.63$ & -- & ...\\
DES\,J235657.00$-$593718.3 & 57220.8 & 359.23748 & -59.62174 & 20.68 & 19.78 & 14.9 & $-32.76 \pm 1.86$ & $6.57 \pm 0.47$ & -- &    0\\
... & 57312.2 & ... & ... & ... & ... &  5.5 & $-28.15 \pm 2.81$ & $7.53 \pm 1.26$ & -- & ...\\
... & 57630.9 & ... & ... & ... & ... & 26.5 & $-34.04 \pm 1.21$ & $6.65 \pm 0.40$ & -- & ...\\
DES\,J235700.42$-$593043.1 & 57312.2 & 359.25174 & -59.51196 & 19.69 & 19.32 &  6.0 & $221.45 \pm 6.06$ & -- & -- &    0\\
DES\,J235701.15$-$593153.2 & 57220.8 & 359.25480 & -59.53145 & 20.13 & 19.83 & 10.5 & $225.35 \pm 2.60$ & $4.66 \pm 0.42$ & -- &    0\\
DES\,J235703.01$-$593824.6 & 57630.9 & 359.26254 & -59.64017 & 21.57 & 21.34 &  5.2 & $-102.37 \pm 6.41$ & -- & -- &    1\\
DES\,J235704.33$-$593151.9 & 57220.8 & 359.26805 & -59.53107 & 21.00 & 20.14 & 11.8 & $51.37 \pm 2.24$ & $4.49 \pm 0.73$ & -- &    0\\
... & 57630.9 & ... & ... & ... & ... & 20.6 & $50.95 \pm 1.52$ & $6.72 \pm 0.45$ & -- & ...\\
DES\,J235704.40$-$593951.1 & 57312.2 & 359.26833 & -59.66419 & 20.29 & 19.59 &  5.4 & $56.31 \pm 3.12$ & -- & -- &    0\\
DES\,J235704.96$-$593543.4 & 57220.8 & 359.27067 & -59.59540 & 20.99 & 20.19 & 11.1 & $176.95 \pm 1.92$ & $4.45 \pm 0.51$ & -- &    0\\
DES\,J235707.45$-$593742.9 & 57220.8 & 359.28106 & -59.62858 & 19.73 & 19.31 & 18.4 & $-101.71 \pm 1.55$ & $1.32 \pm 0.46$ & $-2.78 \pm 0.40$ &    1\\
... & 57312.2 & ... & ... & ... & ... &  6.3 & $-101.93 \pm 3.68$ & $2.13 \pm 0.60$ & $-2.24 \pm 0.35$ & ...\\
... & 57630.9 & ... & ... & ... & ... & 31.8 & $-102.90 \pm 1.10$ & $1.70 \pm 0.34$ & $-2.50 \pm 0.24$ & ...\\
DES\,J235707.90$-$594223.7 & 57312.2 & 359.28293 & -59.70660 & 20.16 & 19.20 &  8.3 & $-1.97 \pm 2.43$ & $7.13 \pm 0.97$ & -- &    0\\
DES\,J235708.66$-$592723.5 & 57223.3 & 359.28609 & -59.45652 & 18.11 & 17.46 & 15.9 & $31.59 \pm 1.58$ & $5.87 \pm 0.51$ & -- &    0\\
... & 57312.2 & ... & ... & ... & ... & 21.4 & $29.55 \pm 1.33$ & $5.88 \pm 0.43$ & -- & ...\\
DES\,J235709.04$-$593400.3 & 57630.9 & 359.28767 & -59.56675 & 21.82 & 21.52 &  3.9 & $-193.88 \pm 7.22$ & -- & -- &    0\\
DES\,J235710.69$-$593149.9 & 57630.9 & 359.29453 & -59.53053 & 21.04 & 20.85 &  6.8 & $-103.03 \pm 2.15$ & $1.43 \pm 0.43$ & -- &    1\\
DES\,J235712.50$-$593716.4 & 57220.8 & 359.30208 & -59.62123 & 19.33 & 18.84 & 27.8 & $279.01 \pm 1.32$ & $4.63 \pm 0.44$ & -- &    0\\
... & 57223.3 & ... & ... & ... & ... &  5.0 & $269.66 \pm 4.31$ & -- & -- & ...\\
... & 57312.2 & ... & ... & ... & ... &  9.4 & $268.18 \pm 1.65$ & $4.46 \pm 1.23$ & -- & ...\\
... & 57630.9 & ... & ... & ... & ... & 45.5 & $272.18 \pm 1.03$ & $4.69 \pm 0.34$ & -- & ...\\
DES\,J235716.79$-$592851.0 & 57312.2 & 359.31996 & -59.48085 & 19.98 & 19.18 &  7.7 & $112.22 \pm 1.92$ & $4.95 \pm 0.93$ & -- &    0\\
DES\,J235717.09$-$594015.2 & 57223.3 & 359.32120 & -59.67089 & 16.05 & 15.33 & 57.8 & $-46.26 \pm 1.24$ & -- & -- &    0\\
... & 57312.2 & ... & ... & ... & ... & 73.3 & $-44.85 \pm 1.21$ & $5.32 \pm 0.34$ & -- & ...\\
DES\,J235719.02$-$593456.6 & 57220.8 & 359.32925 & -59.58239 & 17.63 & 17.20 & 83.7 & $45.36 \pm 1.21$ & $3.54 \pm 0.33$ & -- &    0\\
DES\,J235721.26$-$593632.4 & 57630.9 & 359.33860 & -59.60901 & 21.35 & 21.15 &  6.0 & $-94.60 \pm 8.99$ & -- & -- &    1\\
DES\,J235722.98$-$593628.5 & 57220.8 & 359.34576 & -59.60793 & 19.84 & 19.21 & 21.6 & $29.07 \pm 1.38$ & $5.29 \pm 0.60$ & -- &    0\\
... & 57630.9 & ... & ... & ... & ... & 36.4 & $27.96 \pm 1.09$ & $5.46 \pm 0.36$ & -- & ...\\
DES\,J235726.03$-$593938.1 & 57220.8 & 359.35845 & -59.66059 & 19.27 & 18.81 & 24.4 & $-98.89 \pm 1.35$ & $1.66 \pm 0.39$ & $-2.62 \pm 0.27$ &    1\\
... & 57223.3 & ... & ... & ... & ... &  5.5 & $-99.65 \pm 5.65$ & -- & -- & ...\\
... & 57312.2 & ... & ... & ... & ... &  9.7 & $-99.82 \pm 1.80$ & $2.44 \pm 0.49$ & $-2.16 \pm 0.27$ & ...\\
... & 57630.9 & ... & ... & ... & ... & 41.7 & $-100.30 \pm 1.06$ & $2.12 \pm 0.34$ & $-2.34 \pm 0.20$ & ...\\
DES\,J235726.62$-$593433.6 & 57630.9 & 359.36092 & -59.57599 & 21.47 & 20.94 &  8.8 & $355.54 \pm 1.88$ & $5.22 \pm 0.58$ & -- &    0\\
DES\,J235728.94$-$593222.6 & 57223.3 & 359.37059 & -59.53962 & 19.04 & 18.63 &  5.9 & $80.31 \pm 3.06$ & -- & -- &    0\\
... & 57312.2 & ... & ... & ... & ... & 10.4 & $73.47 \pm 1.66$ & $4.50 \pm 0.72$ & -- & ...\\
DES\,J235729.11$-$592730.7 & 57223.3 & 359.37128 & -59.45852 & 17.67 & 17.07 & 18.3 & $83.79 \pm 1.49$ & $6.94 \pm 0.57$ & -- &    0\\
... & 57312.2 & ... & ... & ... & ... & 24.5 & $82.62 \pm 1.31$ & $6.32 \pm 0.40$ & -- & ...\\
DES\,J235730.23$-$592930.6 & 57312.2 & 359.37596 & -59.49183 & 19.34 & 18.89 &  7.3 & $-99.51 \pm 1.76$ & $2.43 \pm 0.70$ & $-2.15 \pm 0.38$ &    1\\
DES\,J235730.51$-$593110.4 & 57312.2 & 359.37711 & -59.51955 & 19.34 & 19.08 &  6.7 & $290.69 \pm 2.40$ & $5.70 \pm 0.73$ & -- &    0\\
DES\,J235732.29$-$593913.0 & 57630.9 & 359.38452 & -59.65360 & 20.94 & 20.72 &  6.6 & $-9.24 \pm 1.92$ & -- & -- &    0\\
DES\,J235732.74$-$593453.2 & 57630.9 & 359.38641 & -59.58145 & 21.24 & 21.01 &  6.8 & $-106.67 \pm 4.92$ & $1.10 \pm 0.48$ & -- &    1\\
DES\,J235733.29$-$593545.1 & 57220.8 & 359.38873 & -59.59587 & 17.42 & 16.92 & 95.6 & $-2.55 \pm 1.21$ & $4.83 \pm 0.33$ & -- &    0\\
... & 57223.3 & ... & ... & ... & ... & 21.6 & $-0.63 \pm 1.37$ & $5.34 \pm 0.44$ & -- & ...\\
... & 57312.2 & ... & ... & ... & ... & 31.1 & $-2.18 \pm 1.28$ & $5.47 \pm 0.39$ & -- & ...\\
... & 57630.9 & ... & ... & ... & ... & 143.0 & $-2.27 \pm 1.00$ & $5.32 \pm 0.32$ & -- & ...\\
DES\,J235737.91$-$593723.0 & 57630.9 & 359.40797 & -59.62307 & 21.45 & 21.19 &  5.0 & $145.98 \pm 2.26$ & $4.05 \pm 0.89$ & -- &    0\\
DES\,J235738.48$-$593611.6 & 57220.8 & 359.41034 & -59.60323 & 17.17 & 16.58 & 120.5 & $-102.24 \pm 1.21$ & $2.46 \pm 0.33$ & $-2.60 \pm 0.17$ &    1\\
... & 57223.3 & ... & ... & ... & ... & 29.2 & $-100.30 \pm 1.27$ & $2.86 \pm 0.35$ & $-2.41 \pm 0.17$ & ...\\
... & 57312.2 & ... & ... & ... & ... & 38.7 & $-102.37 \pm 1.22$ & $2.71 \pm 0.35$ & $-2.47 \pm 0.17$ & ...\\
... & 57630.9 & ... & ... & ... & ... & 180.6 & $-101.76 \pm 1.00$ & $2.72 \pm 0.32$ & $-2.47 \pm 0.16$ & ...\\
DES\,J235738.70$-$593650.6 & 57220.8 & 359.41127 & -59.61405 & 19.49 & 18.67 & 34.6 & $34.68 \pm 1.25$ & $5.33 \pm 0.38$ & -- &    0\\
DES\,J235738.96$-$593549.9 & 57630.9 & 359.41235 & -59.59721 & 19.48 & 19.16 & 32.4 & $-18.67 \pm 1.11$ & $5.43 \pm 0.38$ & -- &    0\\
DES\,J235740.73$-$593444.4 & 57220.8 & 359.41969 & -59.57901 & 17.54 & 17.11 & 83.2 & $33.40 \pm 1.21$ & $5.58 \pm 0.33$ & -- &    0\\
... & 57630.9 & ... & ... & ... & ... & 130.6 & $32.25 \pm 1.00$ & $5.06 \pm 0.32$ & -- & ...\\
DES\,J235742.88$-$593509.4 & 57220.8 & 359.42866 & -59.58596 & 18.93 & 18.43 & 38.1 & $95.94 \pm 1.28$ & $4.90 \pm 0.36$ & -- &    0\\
DES\,J235745.45$-$593726.4 & 57632.1 & 359.43936 & -59.62399 & 19.80 & 19.39 &  9.9 & $-99.34 \pm 1.93$ & $1.60 \pm 0.40$ & $-2.56 \pm 0.29$ &    1\\
DES\,J235751.61$-$593233.2 & 57632.1 & 359.46506 & -59.54256 & 18.41 & 17.86 & 22.1 & $42.25 \pm 1.26$ & $5.56 \pm 0.42$ & -- &    0\\
DES\,J235754.88$-$593217.1 & 57632.1 & 359.47868 & -59.53808 & 17.89 & 17.41 & 26.2 & $33.66 \pm 1.11$ & $5.71 \pm 0.38$ & -- &    0\\
DES\,J235755.96$-$593614.7 & 57632.1 & 359.48316 & -59.60410 & 19.92 & 19.10 & 12.6 & $142.29 \pm 1.46$ & $7.05 \pm 0.49$ & -- &    0\\
DES\,J235805.42$-$593630.9 & 57632.1 & 359.52258 & -59.60860 & 19.66 & 19.10 & 11.1 & $-8.11 \pm 1.51$ & $4.90 \pm 0.62$ & -- &    0\\
DES\,J235810.14$-$594410.4 & 57632.1 & 359.54225 & -59.73623 & 19.86 & 19.44 &  8.3 & $183.36 \pm 2.05$ & $4.63 \pm 0.78$ & -- &    0\\
DES\,J235811.57$-$594441.3 & 57632.1 & 359.54819 & -59.74481 & 20.30 & 19.84 &  6.3 & $132.96 \pm 2.18$ & $2.49 \pm 0.42$ & -- &    0\\
DES\,J235811.96$-$594150.0 & 57632.1 & 359.54984 & -59.69723 & 20.00 & 19.75 &  6.2 & $243.32 \pm 2.69$ & -- & -- &    0\\
DES\,J235813.15$-$593721.4 & 57632.1 & 359.55478 & -59.62260 & 19.90 & 19.10 & 12.1 & $59.45 \pm 1.61$ & $4.57 \pm 0.53$ & -- &    0\\
DES\,J235814.12$-$593112.7 & 57632.1 & 359.55884 & -59.52019 & 20.00 & 19.65 &  3.9 & $155.66 \pm 4.23$ & $3.85 \pm 0.80$ & -- &    0\\
DES\,J235815.00$-$593809.9 & 57632.1 & 359.56251 & -59.63607 & 21.08 & 20.69 &  2.5 & $ 2.37 \pm 4.82$ & -- & -- &    0\\
DES\,J235821.27$-$594229.9 & 57632.1 & 359.58862 & -59.70831 & 20.80 & 20.37 &  4.1 & $14.22 \pm 3.78$ & $2.68 \pm 0.67$ & -- &    0\\
DES\,J235823.12$-$593250.7 & 57632.1 & 359.59634 & -59.54740 & 19.03 & 18.51 & 11.4 & $73.69 \pm 1.75$ & $5.98 \pm 0.67$ & -- &    0\\
DES\,J235826.09$-$594056.8 & 57632.1 & 359.60869 & -59.68245 & 19.80 & 19.30 &  9.1 & $45.45 \pm 1.79$ & $4.40 \pm 0.65$ & -- &    0\\
DES\,J235828.48$-$594201.9 & 57632.1 & 359.61866 & -59.70052 & 16.56 & 15.92 & 75.3 & $30.95 \pm 1.01$ & $5.87 \pm 0.33$ & -- &    0\\
DES\,J235828.88$-$593527.0 & 57632.1 & 359.62035 & -59.59082 & 19.40 & 18.77 & 13.9 & $42.48 \pm 1.68$ & $5.78 \pm 0.62$ & -- &    0\\
DES\,J235831.24$-$593957.3 & 57632.1 & 359.63018 & -59.66592 & 19.84 & 19.37 &  8.4 & $294.87 \pm 1.40$ & -- & -- &    0\\
DES\,J235832.47$-$593354.4 & 57632.1 & 359.63529 & -59.56511 & 18.23 & 17.68 & 19.9 & $95.50 \pm 1.27$ & $6.13 \pm 0.43$ & -- &    0\\
DES\,J235841.01$-$594300.3 & 57632.1 & 359.67088 & -59.71676 & 19.57 & 19.20 &  9.4 & $35.42 \pm 1.88$ & $5.25 \pm 0.59$ & -- &    0\\

\enddata
\tablenotetext{a}{Quoted magnitudes represent the weighted-average dereddened PSF magnitude derived from the DES images using SExtractor \citep{dw15b}.}
\end{deluxetable*}